\documentclass[aps,floatfix,showpacs,preprintnumbers,amsmath,amssymb]{revtex4}
\usepackage{natbib}
\usepackage{dcolumn}
\usepackage{graphicx}

\newcommand{\physrep}{Phys.~Rep.}

\newcommand{\be}{\begin{equation}}
\newcommand{\ee}{\end{equation}}
\newcommand{\bea}{\begin{eqnarray}}
\newcommand{\eea}{\end{eqnarray}}

\def\[{\begin{equation}}
\def\]{\end{equation}}
\begin{document}
\title{Constraints on growth index parameters from current and future observations}
\author{Jason Dossett$^1$\footnote{Electronic address: jnd041000@utdallas.edu}, Mustapha Ishak$^1$\footnote{Electronic address: mishak@utdallas.edu}, Jacob Moldenhauer$^1$\footnote{Electronic address: jam042100@utdallas.edu},Yungui Gong$^{2}$\footnote{Electronic address: gongyg@cqupt.edu.cn}, Anzhong Wang$^3$\footnote{Electronic address: anzhong\_wang@baylor.edu}}
\affiliation{
$^1$Department of Physics, The University of Texas at Dallas, Richardson, TX 75083, USA;\\
$^2$College of Mathematics and Physics, Chongqing University of Posts and Telecommunications, Chongqing 400065, China,\\
$^3$CASPER, Physics
Department, Baylor University, Waco, TX 76798, USA}
\date{\today}
\begin{abstract}
We use current and future simulated data of the growth rate of large scale structure in combination with data from supernova, BAO, and CMB surface measurements, in order to put constraints on the growth index parameters. We use a recently proposed parameterization of the growth index that interpolates between a constant value at high redshifts and a form that accounts for redshift dependencies at small redshifts. We also suggest here another exponential parameterization with a similar behaviour. The redshift dependent parametrizations provide a sub-percent precision level to the numerical growth function, for the full redshift range. Using these redshift parameterizations or a constant growth index, we find that current available data from galaxy redshift distortions and Lyman-alpha forests is unable to put significant constraints on any of the growth parameters. For example both $\Lambda$CDM and flat DGP are allowed by current growth data. We use an MCMC analysis to study constraints from future growth data, and simulate pessimistic and moderate scenarios for the uncertainties. In both scenarios, the redshift parameterizations discussed are able to provide significant constraints and rule out models when incorrectly assumed in the analysis. The values taken by the constant part of the parameterizations as well as the redshift slopes are all found to significantly rule out an incorrect background. We also find that, for our pessimistic scenario, an assumed constant growth index over the full redshift range is unable to rule out incorrect models in all cases.  This is due to the fact that the slope acts as a second discriminator at smaller redshifts and therefore provide a significant test to identify the underlying gravity theory.
\end{abstract}
\maketitle
\section{introduction}
Since its discovery over a decade ago \cite{acc1a}, cosmic acceleration stands as one of the most important and challenging problems in all physics, see for example the reviews \cite{reviews} and references therein.
As discussed in these reviews and others, cosmic acceleration can be caused by the presence of a dark energy component in the universe or alternatively a modification of gravity physics (namely General Relativity) at cosmological scales. A significant step toward the understanding of the cause of cosmic acceleration is to be able to distinguish between the two competing alternatives. 

Indeed, one ongoing approach to understand the origin of cosmic acceleration is to constrain the equation of state of dark energy \cite{reviews} while other approaches rely on comparisons of the cosmic expansion history to the growth rate of large scale structure, see the incomplete list  \cite{lue,Aquaviva,gong08b,polarski,linder,Koyama,Koivisto,Daniel,knox,ishak2006,laszlo,Zhang,Hu}. Namely, it was shown in many of these references and others that two gravitational theories can have very degenerate Hubble curves but yet have distinct functions of the growth rate of large scale structure in the universe.

The growth rate can thus be used in order to constrain the underlying gravity theory and there has been much interest in providing a parameterization of the growth factor function with one or two parameters that are distinct and characteristic for a given gravity theory, again see the partial list \cite{lue,Aquaviva,gong08b,polarski,linder,Koyama,Koivisto,Daniel,knox,ishak2006,laszlo,Zhang,Hu}.

In this paper, we compare parameterizations of the growth index as a function of the redshift to current and future observations. We also introduce a new exponential parameterization for the growth parameter which allows us to easily characterize the asymptotic value of the growth index. For current constraints, we use the growth data (mainly from galaxy redshift distortions and Lyman-alpha forests) from \cite{porto,ness,guzzo,colless,tegmark,ross,angela,mcdonald,viel1,viel2}, the Constitution compilation of supernova data sets \cite{Constitution}, baryon acoustic oscillation (BAO) measurement from the Sloan Digital Sky Survey (SDSS) \cite{sdss6}, and the distance to the surface of last scattering of the CMB as
measured from the Wilkinson Microwave Anisotropy Probe 5 yr data
(WMAP5) \cite{WMAP5}. We also explore, using a Monte-Carlo-Markov-Chain (MCMC) analysis, how well future growth data from galaxy redshift distortions and Lyman-alpha forests will be able to constrain the growth index parameters and to rule out incorrectly assumed underlying gravity theories. 
\section{Growth of linear perturbations}
The growth rate of large scale structure is derived from matter density perturbation $\delta=\delta\rho_m/\rho_m$ in the linear regime that satisfies the simple differential equation:
\be
\label{eq:ODE}
\ddot{\delta}+2 H \dot{\delta} -4 \pi G_{eff} \rho_m \delta =0,
\ee
where the effect of modified gravity is introduced via the expression for $G_{eff}$. Equation (\ref{eq:ODE}) can be written in terms of the logarithmic growth factor $f=d \ln{\delta}/d\ln{a}$, for many gravity theories as
\be
\label{eq:grwthfeq1}
f'+f^2+\left(\frac{\dot{H}}{H^2}+2\right)f=\frac{3}{2}\frac{G_{eff}}{G}\Omega_m.
\ee
where $'$ is for $d/d\ln{a}$.
The growth function $f$ was shown to be well approximated by the ansatz \cite{Peebles,Fry,Lightman,wang}:
\be
f=\Omega_m^\gamma
\ee
where $\gamma$ is the growth index parameter. Reference \cite{Peebles} proposed the approximation
$f(z=0)=\Omega_0^{0.6}$ for matter dominated models and then a more accurate approximation $f(z=0)=\Omega_0^{4/7}$ was proposed in references \cite{Fry,Lightman}. The authors of \cite{wang} considered dark energy models with slowly varying equation of state and found and index parameter $\gamma(\Omega_m,w)$. This has been the subject of more recent discussions, see for example \cite{linder07,gong08b}, and also extended to models with non zero curvature in \cite{Gong09} and \cite{Mortonson}.

As we discuss below, since growth data spans  a wide range of redshift and the growth index is a function of the redshift, it is worth exploring parameterization of the growth index as a function of the redshift.  For example, in references \cite{polarski,gannouji} the authors proposed a redshift dependent parameterization of the growth index that was applied for redshifts $0<z<0.5$ and was given as: $\gamma(z)=\gamma_0 +  \gamma'\,z,$ where $\gamma'\equiv \frac{d\gamma}{dz}(z=0)$. This showed already the usefulness of a variable growth index to distinguish between dark energy models and modified gravity models \cite{polarski,gannouji}.
Most recently, reference \cite{PaperI} proposed a parameterization that covers a wider range of redshift and interpolates to asymptotic values where the value becomes almost a constant including redshifts up to at the CMB scale, and we summarize it in the next sub-section.

\subsection{Interpolated redshift parameterization of the growth factor index}

In order to be able to consider constraints from growth data that span over a high redshift range such as for example $z=0-3.8$ growth data provided in \cite{porto,ness,guzzo,colless,tegmark,ross,angela,mcdonald,viel1,viel2}, and also future higher redshifts data and constraints up to the CMB scale, the authors proposed a parameterization, see \cite{PaperI}, that interpolates smoothly between low/intermediate and high-redshift ranges up to the CMB surface. Similar to the interpolation proposed in \cite{WMAP5} for the equation of state of dark energy (see appendix C there), the proposed parameterization for the growth index is as follows:
\be
\gamma(a)=\tilde\gamma(a)\,\,\frac{1}{1+a_{_{ttc}}/a}+\gamma_{_{early}}\,\,\frac{1}{1+a/a_{_{ttc}}}
\ee
so that $\gamma(a)$ interpolates between the asymptotic $\gamma_{_{early}}$ value at very high redshifts and
\be
\gamma_{_{late}}(a)=\tilde\gamma(a)= \gamma_0 + (1-a) \gamma_a
\ee
at low redshifts.

Similarly, using $a=1/(1+z)$, the parameterization reads
\be
\label{eq:GammaZ}
\gamma(z)=\tilde\gamma(z)\,\,\frac{1}{1+ \frac{1+z\,\,\,\,\,\,\,}{1+z_{_{ttc}}}} +\gamma_{\infty}\,\,\frac{1}{1+ \frac{1+z_{_{ttc}}}{1+z\,\,\,\,\,\,\,}}
\ee
and interpolates between $\gamma_{\infty}$ at high redshift ($z>z_{ttc}$) up to the CMB scale and the following form at lower redshifts (i.e. for $z< z_{ttc}$) 
\be
\label{eq:GammaLateZ}
\tilde\gamma(z)= \gamma_{late}(z)= \gamma_0 + \gamma_a\,\,\Big{(} \frac{z}{z+1} \Big{)}
\ee

We note here that the $z_{ttc}$ is the redshift of transition from a varying growth index parameter $\gamma(z)$ to an almost constant $\gamma_{\infty}$ and is not necessarily the same as $z_{t}$ which characterizes the transition from a decelerating cosmic expansion to an accelerating one. 

As discussed in \cite{PaperI}, the growth index for dark energy models (QCDM) using this parameterization takes the form given by equation (\ref{eq:GammaZ}) where $\tilde\gamma(z)$ is given by equation (\ref{eq:GammaLateZ}), and where 
\begin{equation}
\label{eq:wcdm0}
\gamma_\infty^{QCDM}=\frac{3(1-w)}{5-6w}.
\end{equation}

As it should, Eq. (\ref{eq:wcdm0}) reduces to the well-known $\gamma_\infty^{^{\Lambda CDM}}=\frac{6}{11}$ for the $\Lambda$CDM model. 

It was found in \cite{PaperI} that the redshift interpolated parameterization of the growth index provides significant improvements over a constant index value. It was found that the $f=\Omega_m^\gamma(z)$ fits the growth function that is integrated numerically from the differential equation to better than $0.0065\%$ for models with constant equations of state and to better than $0.15\%$ for models with variable equations of state.

For the Dvali-Gabadadze-Porrati (DGP) models \cite{dgp}, the interpolated parameterization is given by equations (\ref{eq:GammaZ}) and (\ref{eq:GammaLateZ}) but where the asymptotic value is \cite{linder07,gong08b},
\begin{equation}
\gamma_\infty^{DGP}=\frac{11}{16}
\end{equation}
and as for example derived in \cite{gong08b} from   
\begin{equation}
\label{dgpr}
\gamma=\frac{11}{16}+\frac{7}{5632}(1-\Omega_m).
\end{equation}
The interpolated parameterization was found in \cite{PaperI} to provide a fit to better than $0.04\%$ to the growth function for DGP model with $\Omega_m^0 = 0.27$ . 
\begin{center}
\begin{table}
\begin{tabular}{|c|c|c|}\hline
\multicolumn{3}{|c|}{\bfseries Parameters for various QCDM models.}\\ \hline
$\mathbf{(w_0,w_a)}$&$\mathbf{\gamma_\infty}$&$\mathbf{\gamma_b}$\\ \hline
$(-0.8,0)$&$0.5525$&$0.01124$\\ \hline
$(-1,0)$&$0.5457$&$0.01034$\\ \hline
$(-1.2,0)$&$0.5403$&$0.00983$\\ \hline
$(-0.8,-0.3)$&$0.5460$&$0.01512$\\ \hline
$(-1.2,0.8)$&$0.5579$&$-0.00372$\\ \hline \hline
\multicolumn{3}{|c|}{\bfseries Parameters for some EDE models.}\\ \hline
$\mathbf{(w_0,C)}$&$\mathbf{\gamma_\infty}$&$\mathbf{\gamma_b}$\\ \hline
$(-0.972,1.858)$&$0.5522$&$0.00455$\\ \hline \hline
\multicolumn{3}{|c|}{\bfseries Parameters for various DGP models.}\\ \hline 
$\mathbf{\Omega_m^0}$&$\mathbf{\gamma_\infty}$&$\mathbf{\gamma_b}$\\ \hline
$0.21$&$0.6866$&$-0.02815$\\ \hline
$0.262$&$0.6873$&$-0.02314$\\ \hline
$0.31$&$0.6876$&$-0.01920$\\ \hline
\end{tabular}
\caption{\label{table:Expfit}
Best fit values for the exponential parameterization to the growth rate function as generated from the growth differential equation corresponding to quintessence dark energy models (QCDM) with constant and variable equations of state, early dark energy models (EDE) (see for example \cite{Xia}), and DGP models.
}
\end{table}
\end{center}

\subsection{Exponential Parameterization for the growth index}
\begin{figure}
\begin{center}
\begin{tabular}{|c|c|}
\hline
{\includegraphics[width=2.8in,height=2.0in,angle=0]{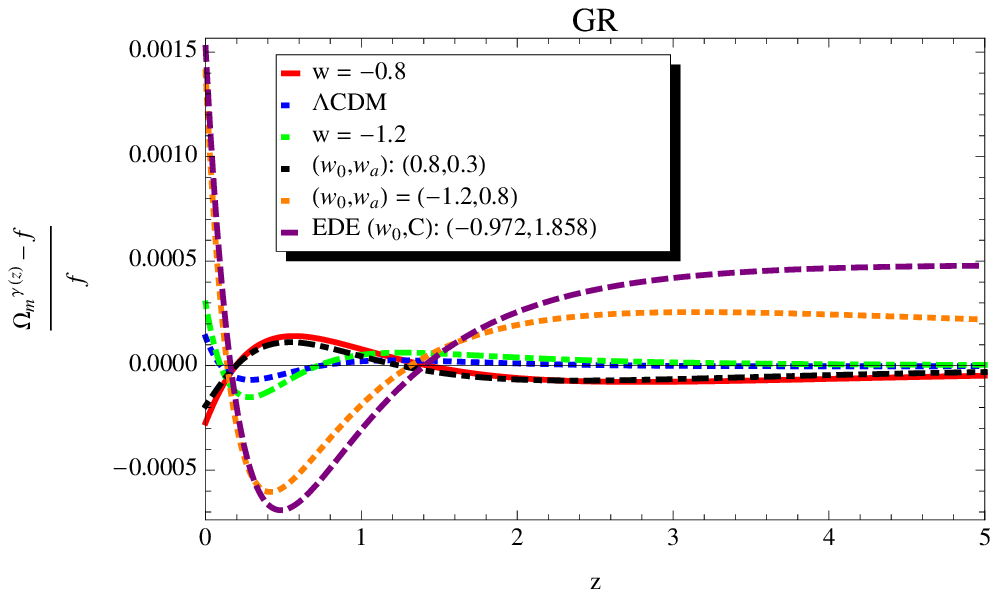}} &
{\includegraphics[width=2.8in,height=2.0in,angle=0]{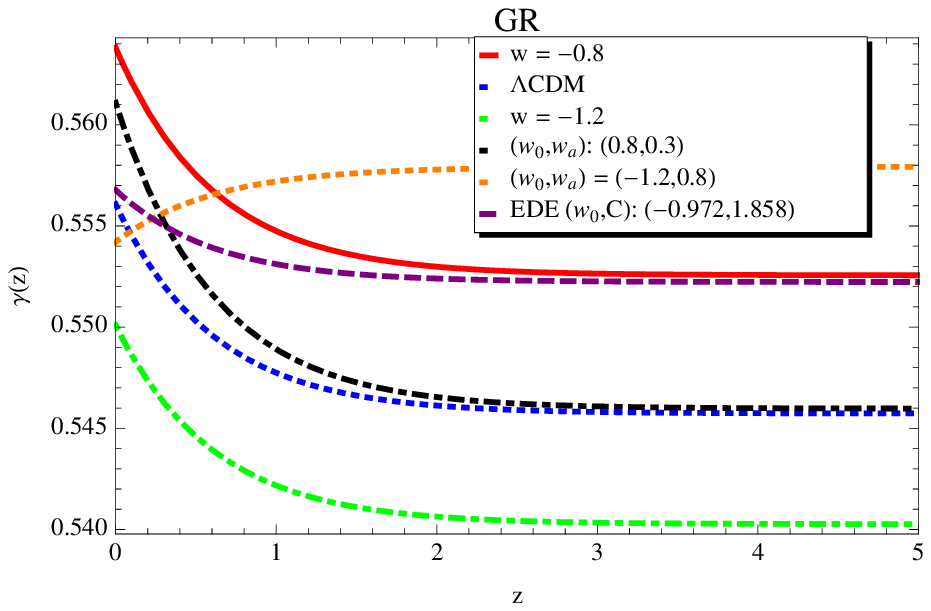}} \\
\hline
{\includegraphics[width=2.8in,height=2.0in,angle=0]{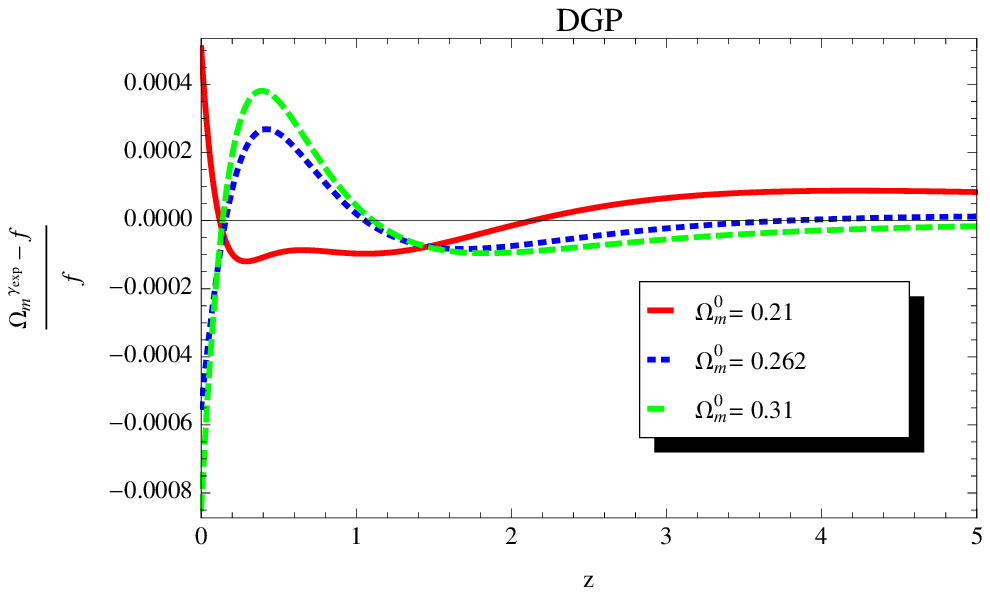}} &
{\includegraphics[width=2.8in,height=2.0in,angle=0]{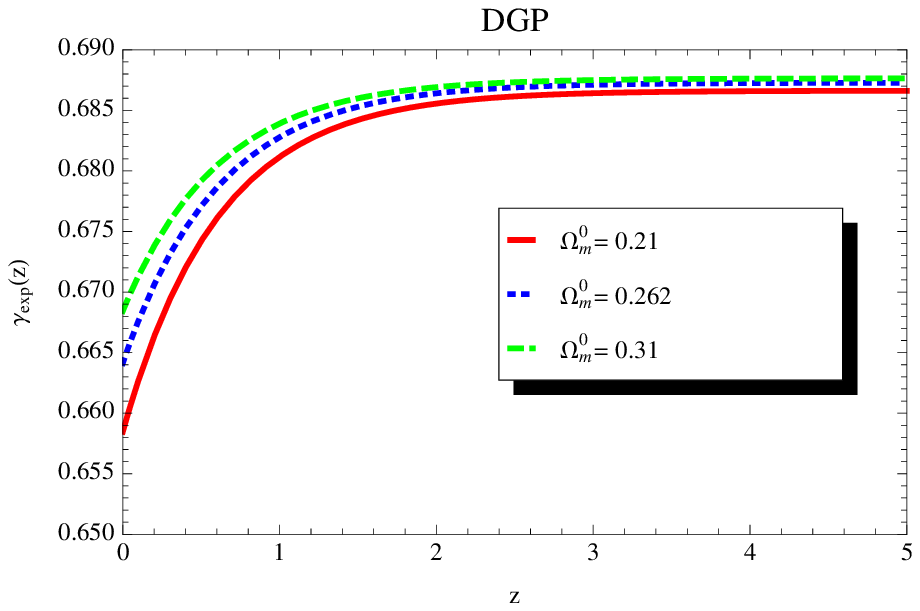}} \\
\hline
\end{tabular}
\caption{\label{fig:CompExp}
Theoretical fittings for the exponential parameterization. TOP-LEFT: We consider various QCDM models and plot the relative error $\frac{\Omega_m^{\gamma(z)}-f}{f}$ in order to compare the fit of the proposed parameterization to that of the growth factor, $f$, that is numerically integrated from the growth ODE. For $\Lambda$CDM we find the best fit parameters $\gamma_{\infty}=0.5457$, $\gamma_b=0.0103$, with $z_{_{ttc}}$ fixed at $0.61$. The fit approximates the growth function $f$ to better than $0.015\%$ while the best fit, $\gamma_{const}^{\Lambda CDM} = 0.5509$ approximates the growth to $0.65\%$ reflecting the discrepancy at low redshifts.
TOP-RIGHT: We plot $\gamma_{exp}(z) = \gamma_{\infty}+\gamma_b (e^{-z/z_{_{ttc}}})$ for various GR models.
BOTTOM-LEFT: We consider the DGP model and plot the relative error $\frac{\Omega_m^{\gamma(z)}-f}{f}$ in order to compare the fit of the proposed parameterization to that of the growth factor $f_{num}$ that is numerically integrated from the growth ODE. We find the best fit parameters $\gamma_{\infty}=0.687$, $\gamma_b=-0.0231$, with fixed $z_{_{ttc}}=0.61$. The fit approximates the growth function $f$ to better than $0.06\%$ for $\Omega_m^0 = 0.262$ while the best fit, constant $\gamma_{const}^{DGP} = 0.6795$ approximates the growth to $2\%$ reflecting the discrepancy at low redshifts.
BOTTOM-RIGHT: We plot $\gamma_{exp}(z) = \gamma_{\infty}+\gamma_b (e^{-z/z_{_{ttc}}})$ for various values of $\Omega_m^0$, finding for all models $\gamma_{\infty} \simeq 0.687$, matching theoretical predictions.
}
\end{center}
\end{figure}
Motivated by the redshift dependence of the growth index, we propose here and explore yet another alternative parameterization for the growth index. We use the fact that the growth index should be nearly constant for early times and vary only during late times. For this reason we introduce the parameterization for the growth index:
\be
\gamma_{exp}(z) = \gamma_{\infty}+\gamma_b (e^{-z/z_{_{ttc}}}).
\label{eq:gamexp}
\ee
Looking at the parameterization one can very easily arrive at the relation $\gamma_{_{z=0}}'= -\gamma_b/z_{ttc}$.  This parameterization is constructed in such a way one can quickly identify $z_{ttc}$ as the redshift at which the growth index begins to transition to the constant value $\gamma_{\infty}$.  It is also worth noting here too, as with the interpolation parameterization above, $z_{ttc}$ is not the same redshift, $z_{t}$, which characterizes the transition from a decelerating cosmic expansion to an accelerating one.  We must also note that our analysis finds that unless there is an extremely small amount of error on the data (on the order of $0.1\%$) it is difficult to constrain this parameter very well. In other words, there is a large range for $z_{ttc}$ for which parametrization provides an excellent fit to the growth. This is similar to the discussion in WMAP paper \cite{WMAP5} about the transition redshift for the equation of state where a large range of redshift was found to fit the data equally well. For our analysis, we fix $z_{ttc}$ to $0.61$ which is found to fit a wide range of models very well.

Fitting parameterization to theoretical data (from numerically integrated growth ODE (\ref{eq:grwthfeq1})) for $\Lambda$CDM and DGP models we find that it is able to fit $\Lambda$CDM to within $0.015\%$ and DGP to within $0.09\%$ for values of $\Omega_{m}^0$ raging from $0.21\mbox{ to }0.31$.  It also shows the ability to correctly pick out the theoretical value of $\gamma_{\infty}$ for both DGP and $\Lambda$CDM to within $0.13\%$ and $0.05\%$, respectively. The results of the fits are provided in Figure \ref{fig:CompExp} and Table \ref{table:Expfit}.

\section{observational constraints}

\subsection{Current data}

In this section, we will use the parameterizations (\ref{eq:GammaZ}) and (\ref{eq:gamexp}) to fit available observational data considering QCDM models and the DGP model. We perform the best fit of the parameters in the various models by $\chi^2$ minimization, ie. minimizing
$\chi^2=\chi^2_{sn}+\chi^2_{bao}+\chi^2_{cmb}$ via a maximum likelihood analysis. In this work, for Type Ia SNe data, we use the 397 Type Ia SNe that compose the Constitution compilation  \cite{Constitution} . This collection of samples consists of 250 high-$z$ Type Ia SNe and 57 nearby Type Ia SNe in the Union set, which were collected by multiple experiments including the ESSENCE Survey \cite{riess,essence}, Supernova Legacy Survey \cite{astier}, an extended dataset of distant SNe observed with the Hubble space telescope, and older observed SNe data, \cite{union} as well as 90 Type Ia SNe from the CFA3 survey \cite{Constitution}. For Type Ia SNe
data, we define the $\chi^2$ for fitting is as
\begin{equation}
\label{chi}
\chi^2_{sn}=\sum_{i=1}^{397}\frac{[\mu_{obs}(z_i)-\mu(z_i)]^2}{\sigma^2_i},
\end{equation}
where $\mu(z)=5\log_{10}[d_L(z)/{\rm Mpc}]+25$ is the extinction-corrected distance modulus, the total uncertainty in the SNe data is $\sigma_i$.  We note that we marginalize over the absolute SNe magnitude, $\mathcal{M}$, when performing our $\chi^2$ minimization.   The luminosity distance is given by the usual relation:
\begin{equation}
\label{lum}
d_L(z)=\frac{1+z}{H_0\sqrt{|\Omega_{k}|}} {\mathcal{S}}\left[\sqrt{|\Omega_{k}|}\int_0^z
\frac{dz'}{E(z')}\right],
\end{equation}
where
\begin{equation}
{\mathcal{S}}(\sqrt{|\Omega_k|}x)=\begin{cases}
\sin(\sqrt{|\Omega_k|}x),& {\rm if}\ \Omega_k<0,\\
\sqrt{|\Omega_k|}x, & {\rm if}\  \Omega_k=0, \\
\sinh(\sqrt{|\Omega_k|}x) & {\rm if}\  \Omega_k>0,
\end{cases}
\end{equation}
and the dimensionless Hubble parameter for dark energy with an equation of state parameterized as $w(z) = w_0+w_a[z/(1+z)]$ is given as
$E(z)=H(z)/H_0=\{\Omega_0(1+z)^3+(1-\Omega_0)[(1+z)^{3(1+w_0+w_a)}\exp(-3 w_a z/(1+z))]\}^{1/2}$
and for the spatially flat DGP model it is:
$E(z)=(1-\Omega_0)/2+[\Omega_0(1+z)^3+(1-\Omega_0)^2/4]^{1/2}$. For the $\Lambda$CDM model, we use $(w_0,w_a)=(-1,0)$.
\begin{table}[tp]
\begin{tabular}{|lcr|}
\hline
\ \ \ \ \ \ \ \ \   $z$\ \ \ \ \ \ \ \ \   & \ \ \ \ \ \ \  $f_{obs}$\ \ \ \ \ \ \  &\ \ References \\
\hline
$0.15$ & $0.49\pm 0.1$ & \cite{colless,guzzo} \\
$0.35$ & $0.7\pm 0.18$ & \cite{tegmark} \\
$0.55$ & $0.75\pm 0.18$ & \cite{ross} \\
$0.77$ & $0.91\pm 0.36$ & \cite{guzzo} \\
$1.4$ & $0.9\pm 0.24$ & \cite{angela} \\
$3.0$ & $1.46\pm 0.29$ & \cite{mcdonald} \\
$2.125-2.72$ & $0.74\pm 0.24$ & \cite{viel1} \\
$2.2-3$ & $0.99\pm 1.16$ & \cite{viel2}\\
$2.4-3.2$ & $1.13\pm 1.07$ & \cite{viel2}\\
$2.6-3.4$ & $1.66\pm 1.35$ & \cite{viel2} \\
$2.8-3.6$ & $1.43\pm 1.34$ & \cite{viel2} \\
$3-3.8$ & $1.3\pm 1.5$ & \cite{viel2}\\
\hline
\end{tabular}
\caption{Summary of observational data used for the growth factor
$f$ from galaxy redshift distortions and Lyman-Alpha forests as compiled in references \cite{porto,ness,guzzo} from original references provided to the right of the table.}
\label{fzdata}
\end{table}

\begin{center}
\begin{table}[t]
\begin{tabular}{|c|c|c|c|c|c|}\hline
\multicolumn{6}{|c|}{\bfseries Constraints on interpolated parameterization for the growth}\\ 
\multicolumn{6}{|c|}{\bfseries index, Eq. (\ref{eq:GammaZ}), using current data}\\ \hline
\multicolumn{1}{|c|}{Model}&\multicolumn{4}{|c|}{Best fit parameters}&\\ \hline
 &\ \ \ \ \ \ \ \ $\mathbf{\gamma_0}$\ \ \ \ \ \ \ \ &\ \ \ \ \ \ \ \ $\mathbf{\gamma_a}$\ \ \ \ \ \ \ \ &\ \ \ \ \ \ \ $\mathbf{\Omega_m^0}$\ \ \ \ \ \ \ &\ \ \ $\mathbf{H_0}$\ \ \ &$\chi^2/dof$\\ \hline 
LCDM&$0.92_{-1.26}^{+1.56}$&$-1.49_{-6.08}^{+6.86}$&$0.251_{-0.023}^{+0.024}$&$72.3_{-2.3}^{+2.5}$&$1.19$\\ \hline
DGP&$0.89_{-1.02}^{+1.49}$&$-2.47_{-4.90}^{+4.88}$&$0.265_{-0.026}^{+0.024}$&$65.5_{-1.9}^{+2.3}$&$1.29$\\ \hline 
\hline
\multicolumn{6}{|c|}{\bfseries Constraints on exponential parameterization for the growth}\\ 
\multicolumn{6}{|c|}{\bfseries index, Eq. (\ref{eq:gamexp}), using current data}\\ \hline
\multicolumn{1}{|c|}{Model}&\multicolumn{4}{|c|}{Best fit parameters}&\\ \hline
 &$\mathbf{\gamma_\infty}$&$\mathbf{\gamma_b}$&$\mathbf{\Omega_m^0}$&$\mathbf{H_0}$&$\chi^2/dof$\\ \hline 
LCDM&$0.16_{-1.11}^{+1.43}$&$0.59_{-2.06}^{+1.72}$&$0.249_{-0.022}^{+0.027}$&$72.5_{-2.3}^{+2.3}$&$1.19$\\ \hline
DGP&$0.03_{-0.77}^{+0.88}$&$0.82_{-1.34}^{+1.26}$&$0.259_{-0.019}^{+0.030}$&$66.0_{-2.1}^{+1.6}$&$1.30$\\ \hline 
\hline
\multicolumn{6}{|c|}{\bfseries Constraints on a constant growth index, $\gamma$, using current data}\\ \hline
\multicolumn{1}{|c|}{Model}&\multicolumn{2}{|c|}{$\mathbf{\gamma}$}&$\mathbf{\Omega_m^0}$&$\mathbf{H_0}$&$\chi^2/dof$\\ \hline
LCDM&\multicolumn{2}{|c|}{$0.60_{-0.28}^{+0.35}$}&$0.250_{-0.020}^{+0.023}$&$72.3_{-2.0}^{+2.3}$&$1.19$\\ \hline
DGP&\multicolumn{2}{|c|}{$0.54_{-0.24}^{+0.30}$}&$0.262_{-0.022}^{+0.023}$&$65.6_{-1.7}^{+1.9}$&$1.30$\\ \hline 
\end{tabular}
\caption{\label{table:obs}
Constraints from the combined current observational data including the growth data given in TABLE \ref{fzdata} on the growth index parameters, $\Omega_m^0$, and $H_0$. We see that current data is unable to constrain any of the growth index parameters well enough to draw any conclusions.   
}
\end{table}
\end{center}

To get constraints from observations of the CMB we follow \cite{WMAP5} and define three fitting parameters for comparison to WMAP5 data: the shift parameter, $R$, 
\be
R(z_*)=\sqrt{\Omega_m}H_0(1+z_*)D_A(z_*),
\label{eq:ShiftParameter}
\ee
with the redshift of the surface of last scattering, $z_*$, given by: 
\be
z_*=1048[1+0.00124(\Omega_b h^2)^{-0.738}][1+g_1 (\Omega_m h^2)^{g_2}].
\label{eq:zstar}
\ee
The constants $g_1$ and $g_2$ in the above expression are:
\be
g_1=\frac{0.0783(\Omega_b h^2)^{-0.238}}{1+39.5(\Omega_b h^2)^{0.763}},
\label{eq:zstarg1}
\ee
and
\be
g_2=\frac{0.560}{1+21.1(\Omega_b h^2)^{1.81}},
\label{eq:ztarg2}
\ee
A third parameter, the acoustic scale, $l_a$, is
\be
l_a=(1+z_*)\frac{\pi D_A(z_*)}{r_s(z_*)},
\label{eq:AcousticScale}
\ee
with the proper angular diameter distance, $D_A(z)=D_L(z)/(1+z)^2$ and the comoving sound horizon, $r_s(z_*)$ 
\be
r_s(z_*)=\frac{1}{\sqrt{3}}\int^{1/(1+z_*)}_{0}{\frac{da}{a^2H(a)\sqrt{1+(3\Omega_b/4\Omega_{\gamma})a}}},
\label{eq:SoundHorizon}
\ee
where $\Omega_{\gamma}=2.469\times 10^{-5}h^{-2}$ for $T_{cmb}=2.725 K$.

Together the parameters $x_i=(R,l_a,z_*)$ are used to fit $\chi^2_{CMB}=\triangle x_iCov^{-1}(x_ix_j)\triangle x_j$ with $\triangle x_i=x_i-x^{obs}_i$ and $Cov^{-1}(x_ix_j)$is the inverse covariance matrix for the parameters. 

Next, for the BAO, we follow \cite{sdss6} and define the ratio of the sound horizon, $r_s(z_d)$ to the effective distance, $D_V$ as a fit for SDSS by
\be
\chi^2_{BAO}=\Big(\frac{r_s(z_d)/D_V(z=0.2)-0.198}{0.0058}\Big)^2+\Big(\frac{r_s(z_d)/D_V(z=0.35)-0.1094}{0.0033}\Big)^2,
\label{eq:chisqbao}
\ee
with 
\be
D_V(z)=\Big(D_A^2(z)(1+z)^2\frac{z}{H(z)}\Big)^{1/3},
\label{eq:EffectiveDistance}
\ee
and the redshift, $z_d$ as
\be
z_d=\frac{1291(\Omega_m h^2)^{0.251}}{1+0.659(\Omega_m h^2)^{0.828}}[1+b_1(\Omega_b h^2)^{b_2}],
\label{eq:zdrag}
\ee
where
\be
b_1=0.313(\Omega_m h^2)^{-0.419}[1+0.607(\Omega_m h^2)^{0.674}],
\label{eq:zdragb1}
\ee
and
\be
b_2=0.238(\Omega_m h^2)^{0.223}.
\label{eq:zdragb2}
\ee

We also add the prior $H_0=72\pm 8$ km/s/Mpc given by \cite{freedman}. 

For the growth factor data, we define
\begin{equation}
\label{fzchi}
\chi^2_f=\sum_{i=1}^{12}\frac{[f_{obs}(z_i)-\Omega^\gamma(z_i)]^2}{\sigma_{fi}^2},
\end{equation}
where $\sigma_{fi}$ is the $1\sigma$ uncertainty in the $f(z)$ data.

The growth data that we use (Table \ref{fzdata}) is data that was converted 
in the work of \cite{porto,ness,guzzo} from either measurement of redshift distorsion 
parameter $\beta=f/b$ (where $b$ is the bias that measures how galaxies trace the 
mass density field) or from various power spectrum amplitudes from Lyman-$\alpha$ Forest data. 
The list of the respective original references is also given in the table. A caveat in 
using this data to constrain other cosmological models is that at various 
steps in the process of analysing or converting the data (in the references above and 
those in Table \ref{fzdata} ), the $\Lambda$CDM model was assumed. 
So if one wants to use the data to constrain other models, and in particular 
modified gravity models, one than should redo all the steps assuming that 
model, starting from original observations. Therefore one should keep 
in mind this caveat when looking at the results below for the DGP model. 

Monte Carlo Markov Chains (MCMC's) are used to compute the likelihoods for 
the parameters in the model.  This method randomly chooses values for the above parameters and, 
based on the $\chi^2$ obtained, either accepts or rejects the set of parameters 
via the Metropolis-Hastings algorithm.  When a set of parameters is accepted it 
is added to the chain and forms a new starting point for the next step.  
The process then repeated until the specified convergence is reached. In 
this work the MCMC code we use is based on a modified version of the publicly 
available package CosmoMC \cite{cosmomc}.  
\begin{figure}[t]
\begin{center}
\begin{tabular}{|c|c|}
\hline
{\includegraphics[width=2.8in,height=2.0in,angle=0]{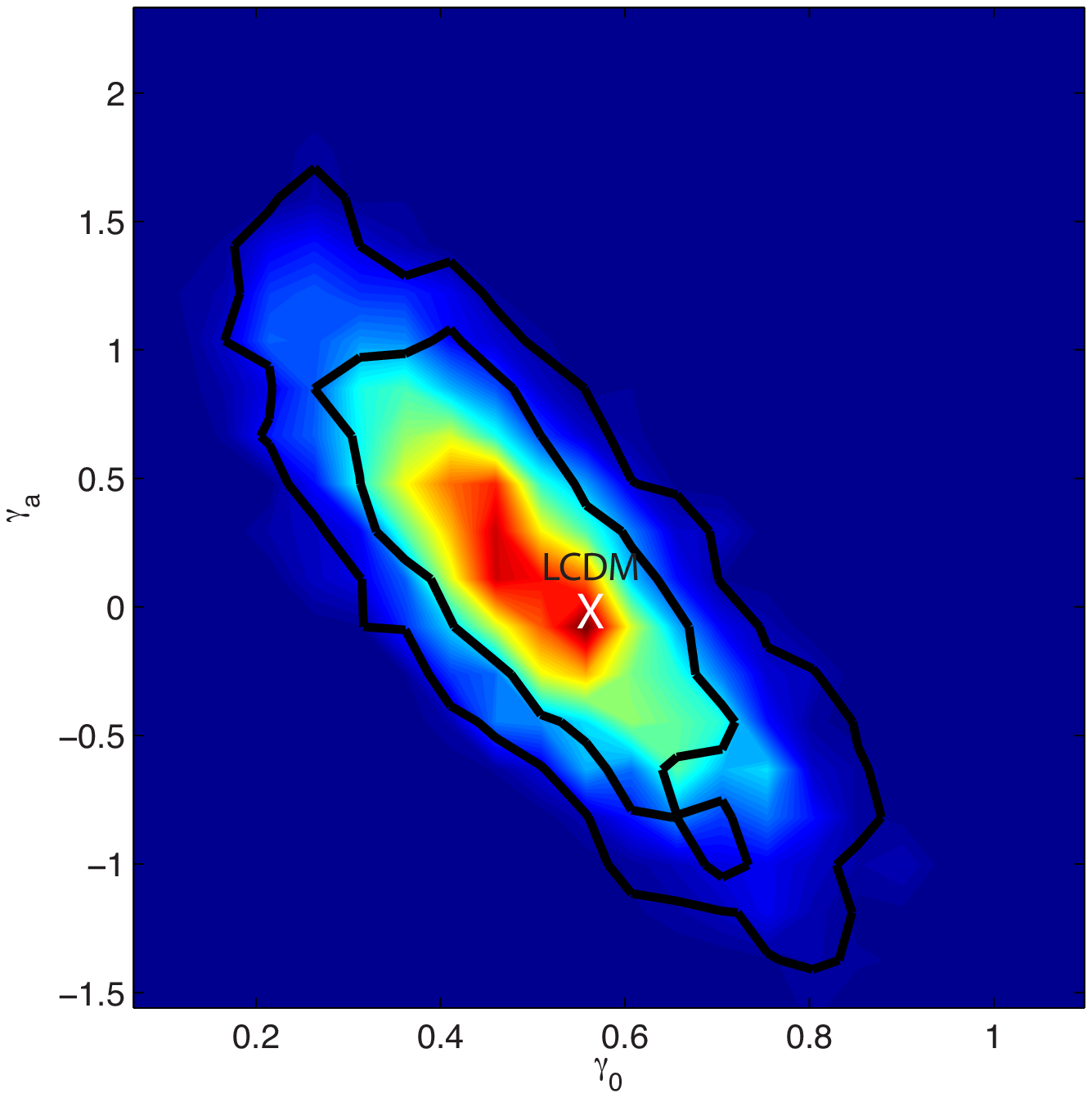}} &
{\includegraphics[width=2.8in,height=2.0in,angle=0]{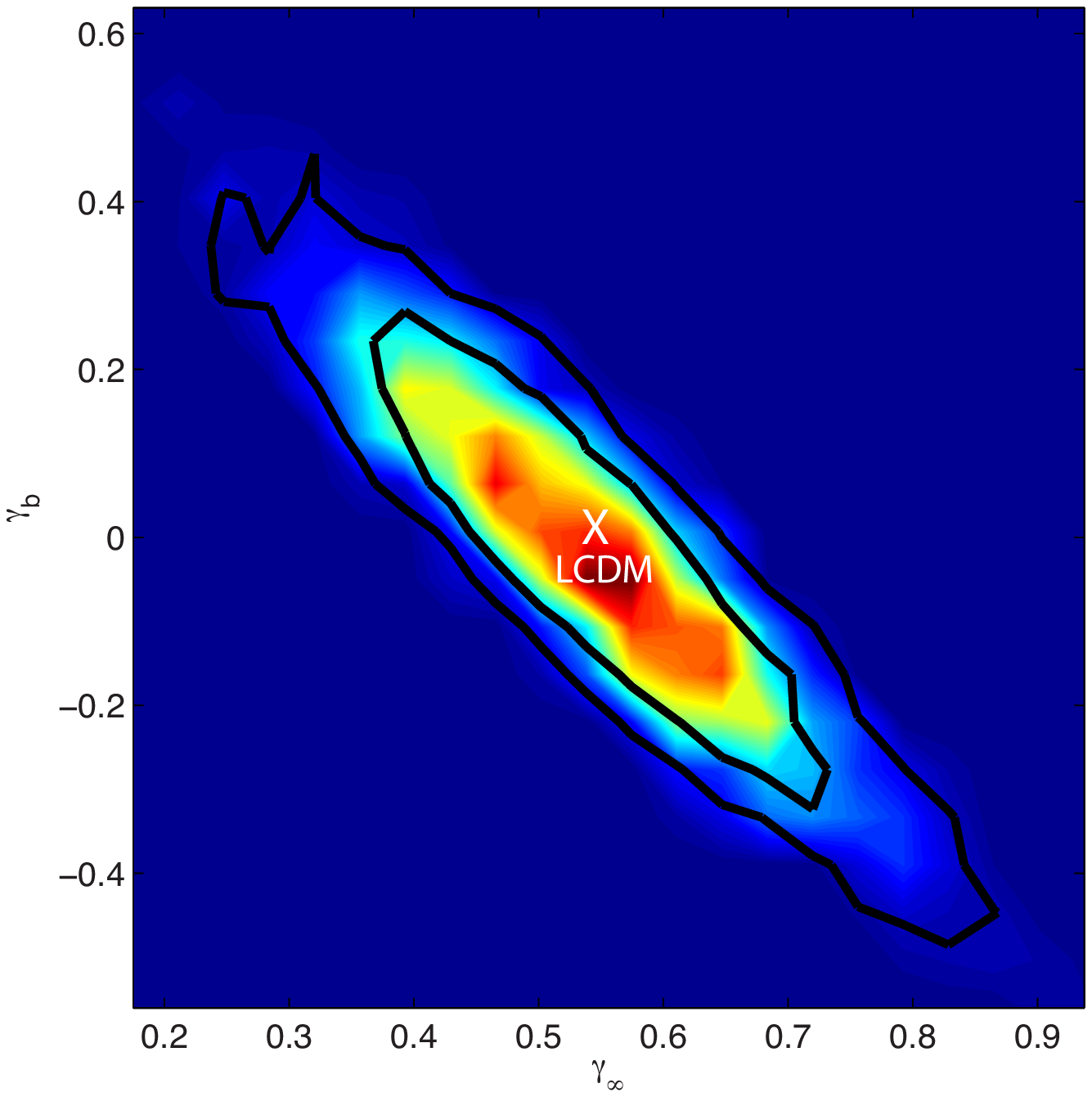}} \\
\hline
\end{tabular}
\caption{\label{fig:LCDMonLCDM}
As discussed in section III-B, this is a first simple test to check (using future simulated data) the working of the best fit method for the growth index parameters using our MCMC analysis. The growth data is generated using the moderate scenario using $\Lambda$CDM model as the background  model as well as the fitted theoretical model. 
TOP LEFT: Interpolated parameterization: $\chi^2/dof = 0.854$. TOP RIGHT: Exponential parametrization: $\chi^2/dof = 0.855$. As expected, for each parametrization, we recover best fit parameters of the $\Lambda$CDM model that are practically the same as the ones found from other fits in \cite{PaperI} for the interpolated parametrization, and in Figure \ref{fig:CompExp} of this paper for the exponential parametrization.  We additionally recover the best fit values for $\Omega_m^0$ and $H_0$ that deviate less than $0.4\%$ from the values used to generate the simulated future data.
}
\end{center}
\end{figure}

We perform the fits using $\Lambda$CDM and flat DGP with all the combined 
data above and allow the following concerned parameters to vary: the physical dark matter density, $\Omega_{dm}h^2$; the physical baryon density,
$\Omega_{b}h^2$; the ratio of the sound horizon to the angular
diameter distance, $\theta$ ($\Omega_m^0$ and $H_0$ are derived parameters in CosmoMC); and the growth parameters. The values found for the growth parameters as well as $\Omega_m^0$ and $H_0$ are summarized in Table \ref{table:obs}. We find that current data is unable to put significant constraints on any forms of the growth factor index parameters and we turn to future simulated data.

\subsection{Simulated future data}
\begin{figure}[t]
\begin{center}
\begin{tabular}{|c|c|}
\hline
{\includegraphics[width=2.8in,height=2.0in,angle=0]{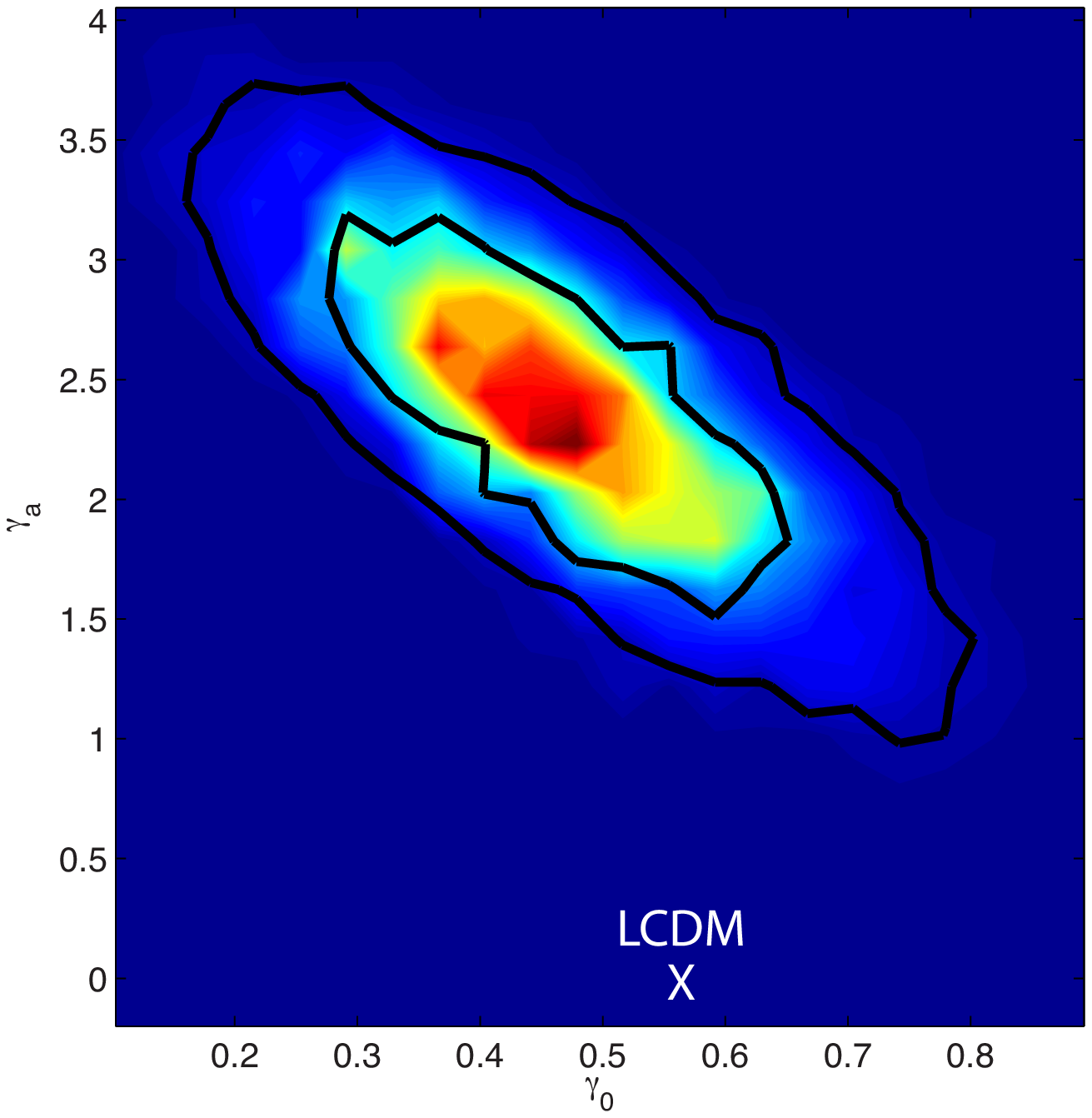}} &
{\includegraphics[width=2.8in,height=2.0in,angle=0]{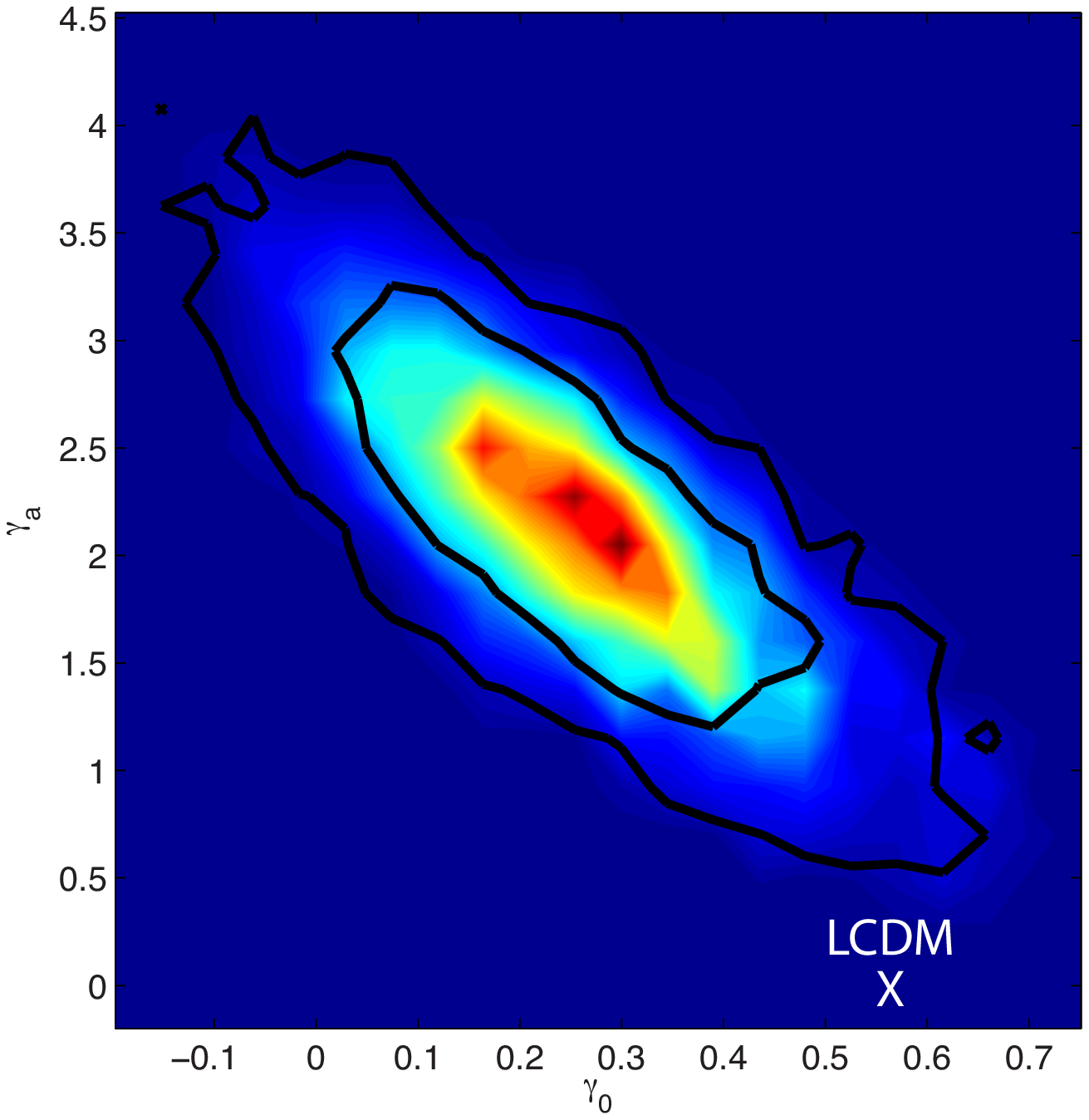}} \\
\hline
{\includegraphics[width=2.8in,height=2.0in,angle=0]{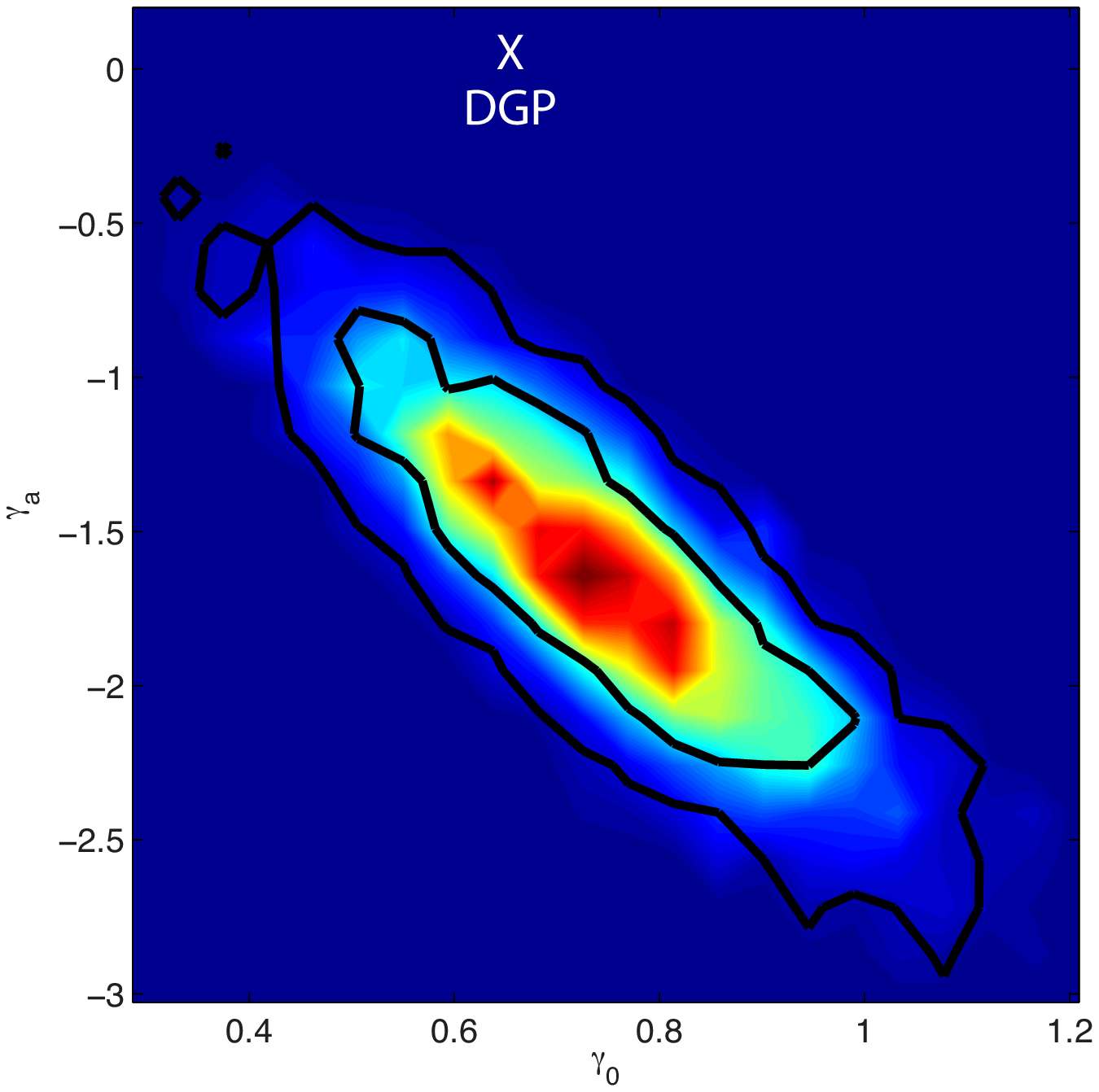}} &
{\includegraphics[width=2.8in,height=2.0in,angle=0]{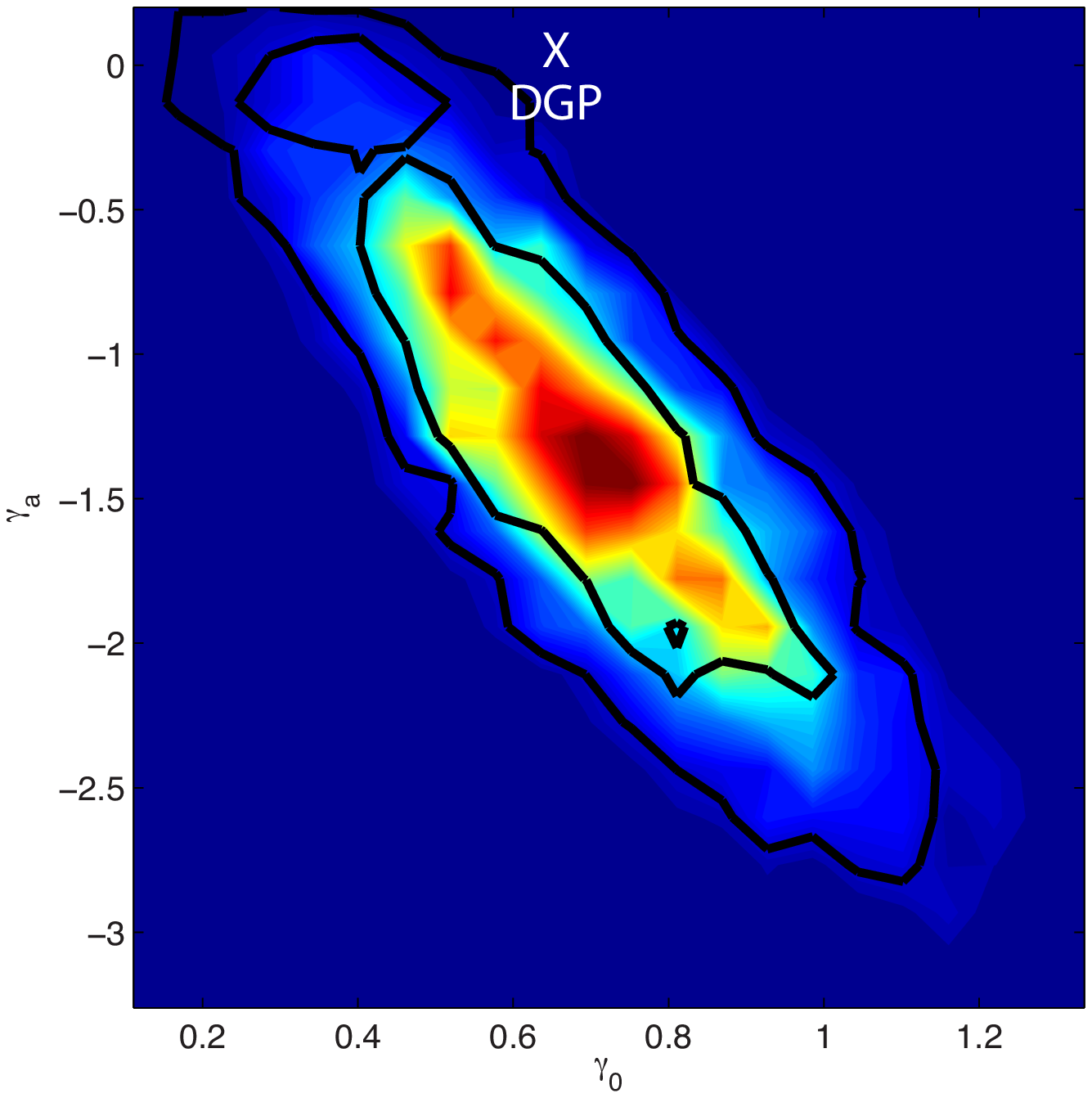}} \\
\hline
\end{tabular}
\caption{\label{fig:INT}
Interpolated parameterization.  TOP LEFT: Moderate scenario fitting
background DGP data to an assumed $\Lambda$CDM theoretical model:
$\chi^2/dof = 1.038$.  TOP RIGHT: Pessimistic scenario fitting background
DGP data to an assumed $\Lambda$CDM theoretical model: $\chi^2/dof = 1.023$.
BOTTOM LEFT:   Moderate scenario fitting background $\Lambda$CDM data to an
assumed DGP theoretical model: $\chi^2/dof = 1.037$.  BOTTOM RIGHT:
Pessimistic scenario fitting background $\Lambda$CDM data to an assumed DGP
theoretical model: $\chi^2/dof = 1.017$. As shown on the figures, in each
case the incorrect assumed background model is ruled out to $95.45\%$.
}
\end{center}
\end{figure}
The next question that one needs to address is how well future data could constrain the growth index parameters. We use here a modified version of the program CosmoMC \cite{cosmomc} to include growth rate codes and likelihood function in order to find the best fit growth index parameters. We simulate  growth data and uncertainties using growth ODEs' codes and use in the analysis pessimistic and moderate scenarios based on values of the one sigma uncertainties on future growth data $f_{obs}$.  We also simulate future data for Type Ia SNe, BAO's and CMB surface measurements and use all the simulated data sets in order to see what constraints future data can put on the growth parameters even when other cosmological parameters are varied. 

To simulate future growth data, we use the discussions and some current data provided in the following references \cite{guzzo,viel1,viel2,Xia,gratton,white} for data coming from redshift distortions or from Lyman-Alpha forests, in order to extrapolate two scenarios. It is expected that the limitations will come from the systematics in each probes. We generate 80 points (or bins) for the growth rate that are almost equally spaced by $\Delta z=0.05$ between redshifts 1 and 4, but where we allow for random small departures from these exact locations up to $1.5\sigma$ were $\sigma$ is the uncertainty on the simulated data point as described below. 

We use a pessimistic scenario where we assume that we will have more data but the uncertainties will get only slightly better then the ones of the current data and we use the one sigma uncertainties on the observed values of the growth as follows: $\sigma(f_{obs})=20\%$ for the range $0<z\le 2.0$ and $\sigma(f_{obs})=30\%$ for $2<z\le 4.0$. The moderate scenario is:  $\sigma(f_{obs})=15\%$ (an improvement of a factor of 1.33) for the range $0<z\le 2.0$ and $\sigma(f_{obs})=25\%$ (an improvement of a factor of 1.2) for $2<z\le 4.0$.

\begin{figure}[t]
\begin{center}
\begin{tabular}{|c|c|}
\hline
{\includegraphics[width=2.8in,height=2.0in,angle=0]{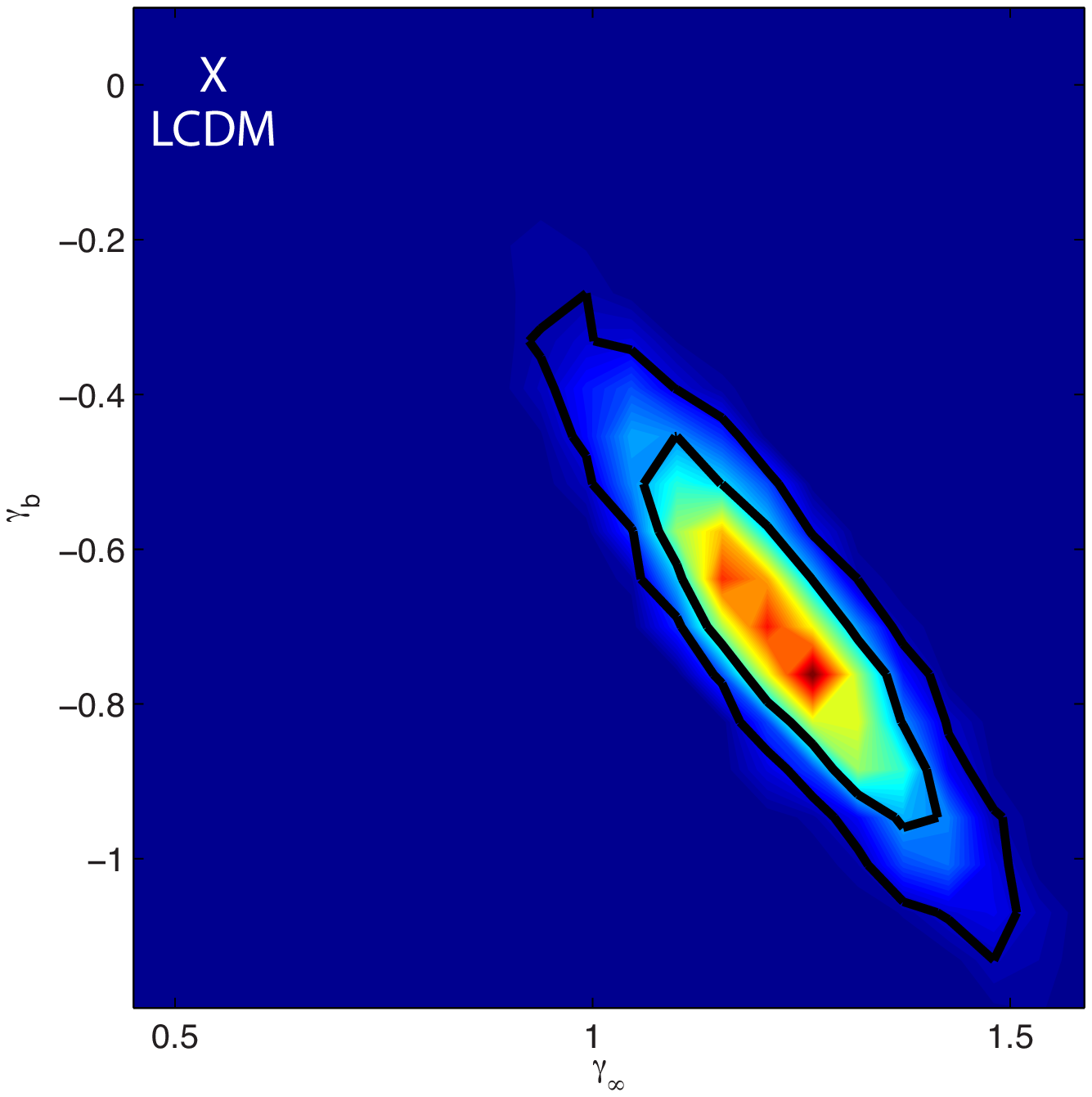}} &
{\includegraphics[width=2.8in,height=2.0in,angle=0]{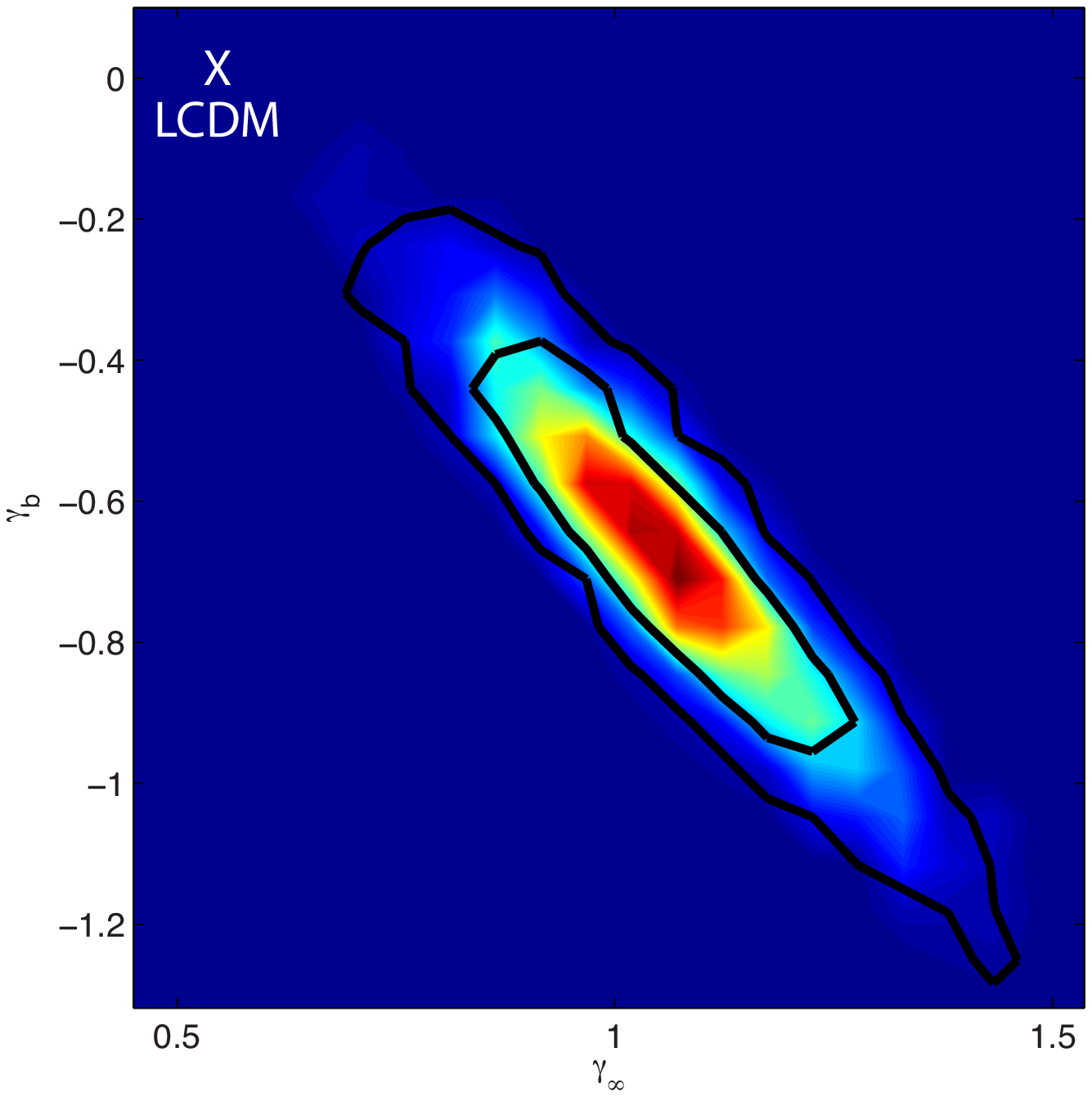}} \\
\hline
{\includegraphics[width=2.8in,height=2.0in,angle=0]{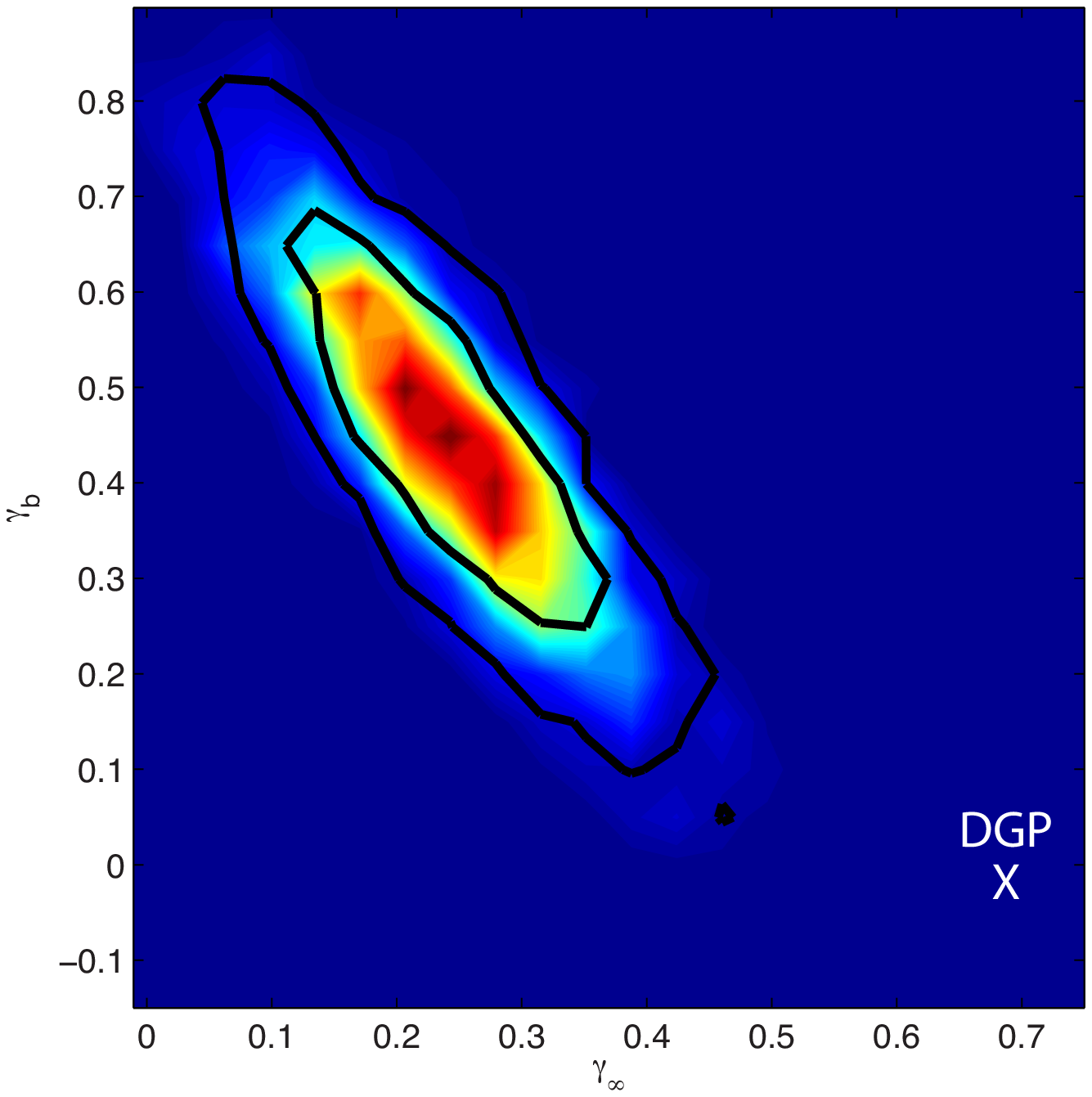}} &
{\includegraphics[width=2.8in,height=2.0in,angle=0]{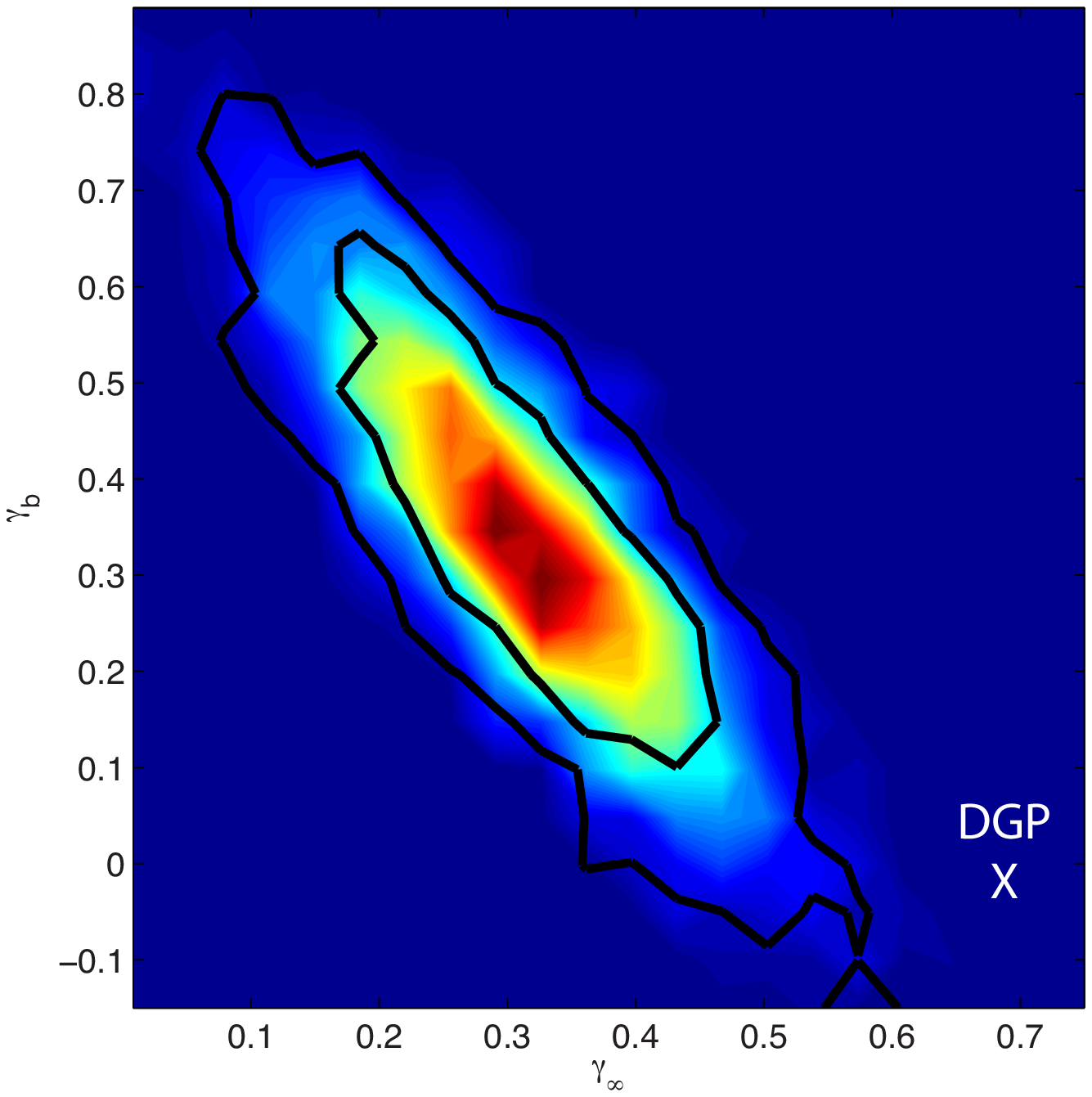}} \\
\hline
\end{tabular}
\caption{\label{fig:Exp}
Exponential parameterization. TOP LEFT: Moderate scenario fitting background DGP data to an assumed $\Lambda$CDM theoretical model: $\chi^2/dof = 1.032$.  TOP RIGHT: Pessimistic scenario fitting background DGP data to an assumed $\Lambda$CDM theoretical model: $\chi^2/dof = 1.022$.  BOTTOM LEFT:   Moderate scenario fitting background $\Lambda$CDM data on an assumed DGP theoretical model: $\chi^2/dof = 1.032$.  BOTTOM RIGHT: Pessimistic scenario fitting background $\Lambda$CDM data to an assumed DGP theoretical model: $\chi^2/dof = 1.015$. As shown on the figures, in each case the incorrect assumed background model is ruled out to $95.45\%$.
}
\end{center}
\end{figure}

We simulate the future data by using our best-fit results for
cosmological parameters from current data to generate 
the cosmological distances parameters for each future data set, i.e. $\mu(z)$,
$f_{obs}$, $r_s(z_d)/D_V$, and $(R,l_a,z_*)$ for supernovae, growth,
BAO, and CMB surface, respectively. For BAO future constraints we used as guidance a Fisher matrix analysis performed in \cite{SeoEisenstein}. As a conservative approach,  we generate seven data points $0.2\lesssim z \le 0.7$ with a spacing of about $\Delta z=0.075$ and an uncertainty of $2\%$ for $z<0.5$ and $7.5\%$ for $0.5<z<0.7$ to use for BAO constraints.  Here too we allow small random departures from the exact data values up to $1.5\sigma$. Our future data is in line with predictions from BOSS documentation\cite{BOSS}.  
For Type Ia SNe we generate 500 equally spaced data points (distance moduli) for $z=0.0036-1.8$ with an uncertainty of $0.7\%$, in-line with uncertainties on these values from future surveys. Again we allow small random departures from the exact values up to $1.5\sigma$. 
We also use the "Bluebook" documentation of the Planck satellite\cite{BlueBookPlanck} to improve the constraints on the CMB surface measurements from WMAP5.  Conservatively, we decrease the uncertainty by half. 

When we do the fits we use all the simulated data sets and allow for the following parameters to vary: $\Omega_{dm}h^2$, $\Omega_{b}h^2$, $\theta$, and the growth parameters corresponding to the parameterization used for $\gamma$, (as usual, $\Omega_m^0$ and $H_0$ are the derived parameters in CosmoMC). 

We call the background model in our analysis the model that we use in order to generate the simulated data. When we simulate the data for that background model, we use all the functions involved to be those of that model. For example, we use the growth equation (\ref{eq:grwthfeq1}) with $H(z)$ and $G_{eff}$ for that model as given explicitly in our previous work \cite{PaperI} for $\Lambda$CDM and DGP models. Similarly, we use the functions of the background model for the supernova luminosity distance and for the angular diameter distances for the BAO and distances to the CMB surface. In that way all the data has been generated consistently for one background model. 

We call the fitted theoretical model in this analysis the model that we use in order to 
fit the data previously simulated using the background model. In other words, the fitted theoretical model is the assumed theoretical model that one will use in order to fit the real data from a real survey. 

Now, one also needs a strategy on how to apply the tests. A trivial exercise is to generate the data using a given background model as the true cosmology ($\Lambda$CDM, let's say) and then run the best fit test for $\gamma(z)$ and other cosmological parameters using the same theoretical model as the one used to generate the data ($\Lambda$CDM here). In this case, one should expect to recover confidence contours where, of course, $\Lambda$CDM should be the best fit (or not too far from it) and the confidence contours sizes will depend on the uncertainties added to the generated data. As shown in our Figure \ref{fig:LCDMonLCDM} this trivial check works very well, recovering best-fit value for growth parameters very close to those we would expect. In this case, when these fits are performed we also find best fit values for $\Omega_m^0$ and $H_0$ that deviate less than $0.4\%$ from the values used to generate the data.  

Of course, a more significant strategy that we adopt here is to explore when is the test able to rule out a given model. For this, we generate the data using a given \textit{background model} ($\Lambda$CDM, let's say) and then apply the best fit test for $\gamma(z)$ and other cosmological parameters using a different \textit{fitted theoretical model}. In this approach, we are testing if the simulated data and uncertainties (i.e. the survey) are able to rule out the growth index parameters of the fitted theoretical model, as it should, or the uncertainties are too large and more precision is needed from the survey.  

In fact, the values of the growth index parameters given by the literature are attained by taking a given model and evaluating growth index parameters within that \textit{same} model and then the characteristic values of for example $\gamma=0.55$ for $\Lambda$CDM and $\gamma=0.68$ for DGP are obtained, see for example \cite{linder07,gong08b}. This approach gives values for growth index parameters that, by their very construction, are model dependent. For this reason one cannot simply fit growth index parameters using an arbitrary model and expect that the values returned will be values indicating what the correct model of gravity is. More clearly, for a given background model  (\textit{i.e.} the simulated or true data), the values of the growth index parameters that one will obtain from the fit will be different depending on what theoretical model is used for the fit. For example, if one is fitting the growth index and assuming the
cosmology is $\Lambda$CDM (theoretical model used) and finds that
$\gamma = 0.55$ is not within the range of uncertainties, it can be
concluded $\Lambda$CDM is not the correct true cosmology.  However, the
value measured cannot be compared to the value $\gamma=0.68$ of the DGP
model because in this example we are using $\Lambda$CDM as our theoretical
cosmology.  In other words, it does not matter if $\gamma=0.68$ falls
within the uncertainties or not because DGP is not the theoretical
model used in this example. The particular measured value of $\gamma$ cannot be used to confirm or rule out a theoretical model other than the one used in the fit. For this reason, the growth index parameters are best used to rule out or confirm an assumed theoretical model. 
\begin{figure}[t]
\begin{center}
\begin{tabular}{|c|c|}
\hline
{\includegraphics[width=2.4in,height=1.5in,angle=0]{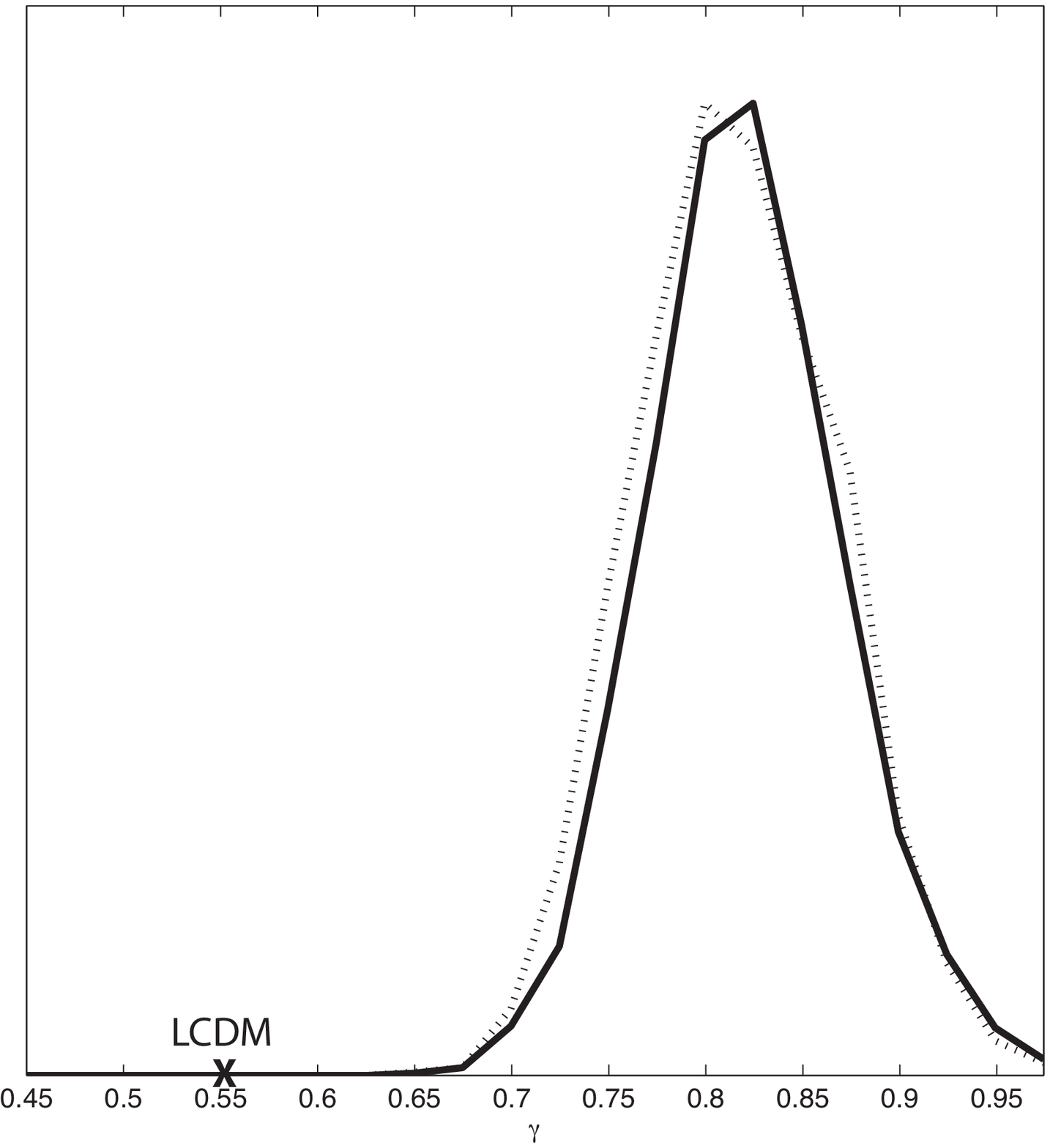}} &
{\includegraphics[width=2.4in,height=1.5in,angle=0]{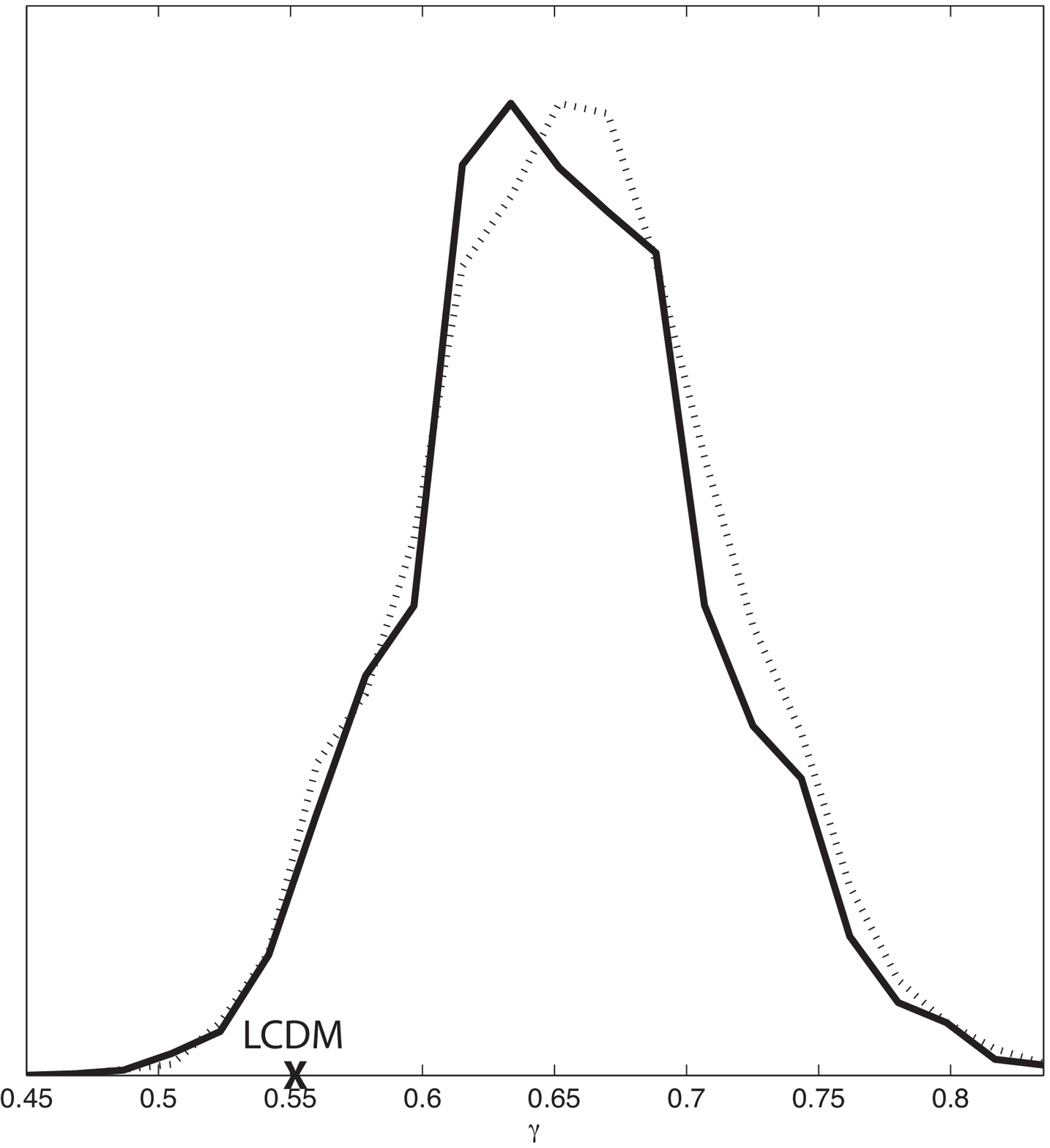}} \\
\hline
{\includegraphics[width=2.4in,height=1.5in,angle=0]{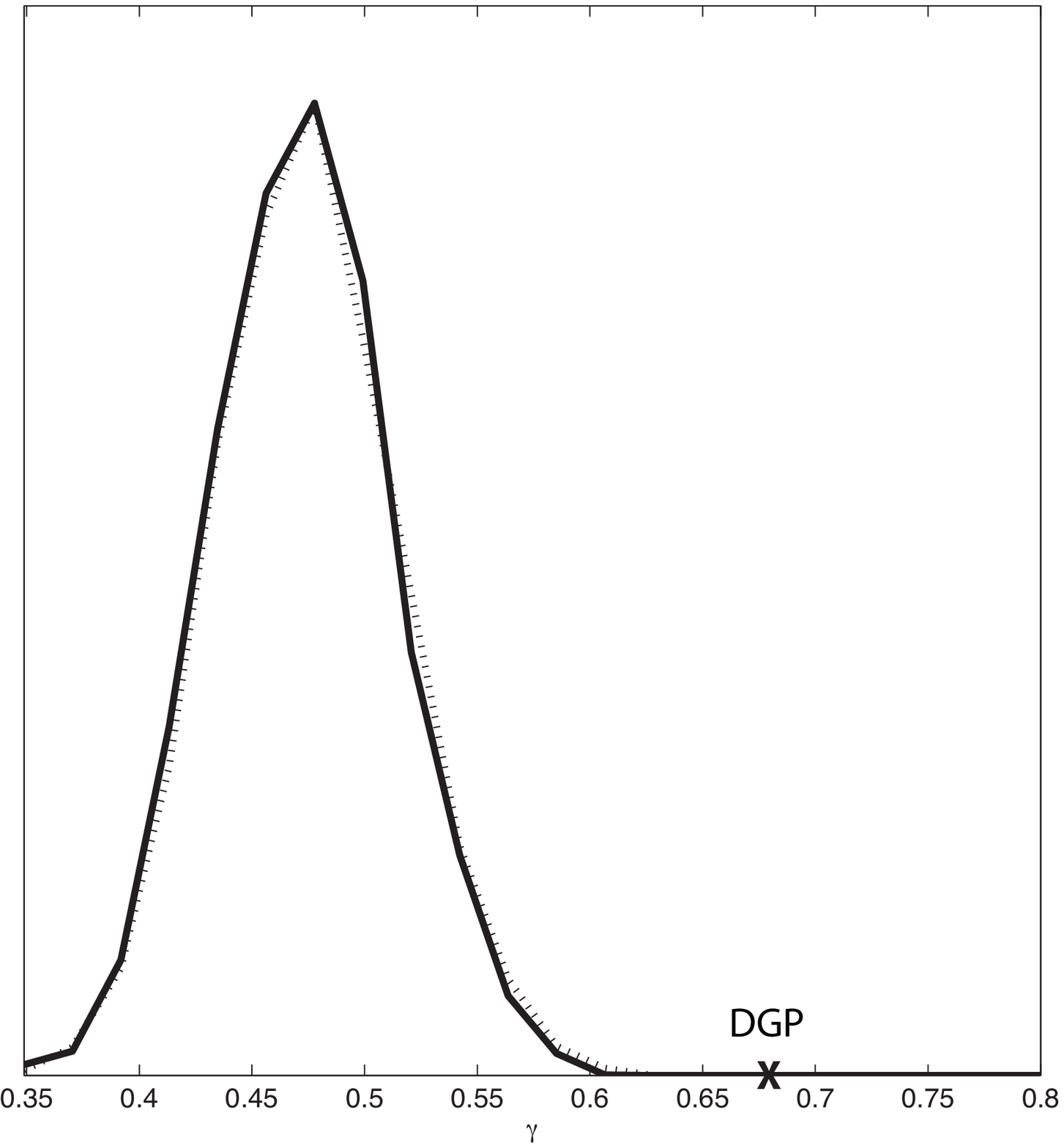}} &
{\includegraphics[width=2.4in,height=1.5in,angle=0]{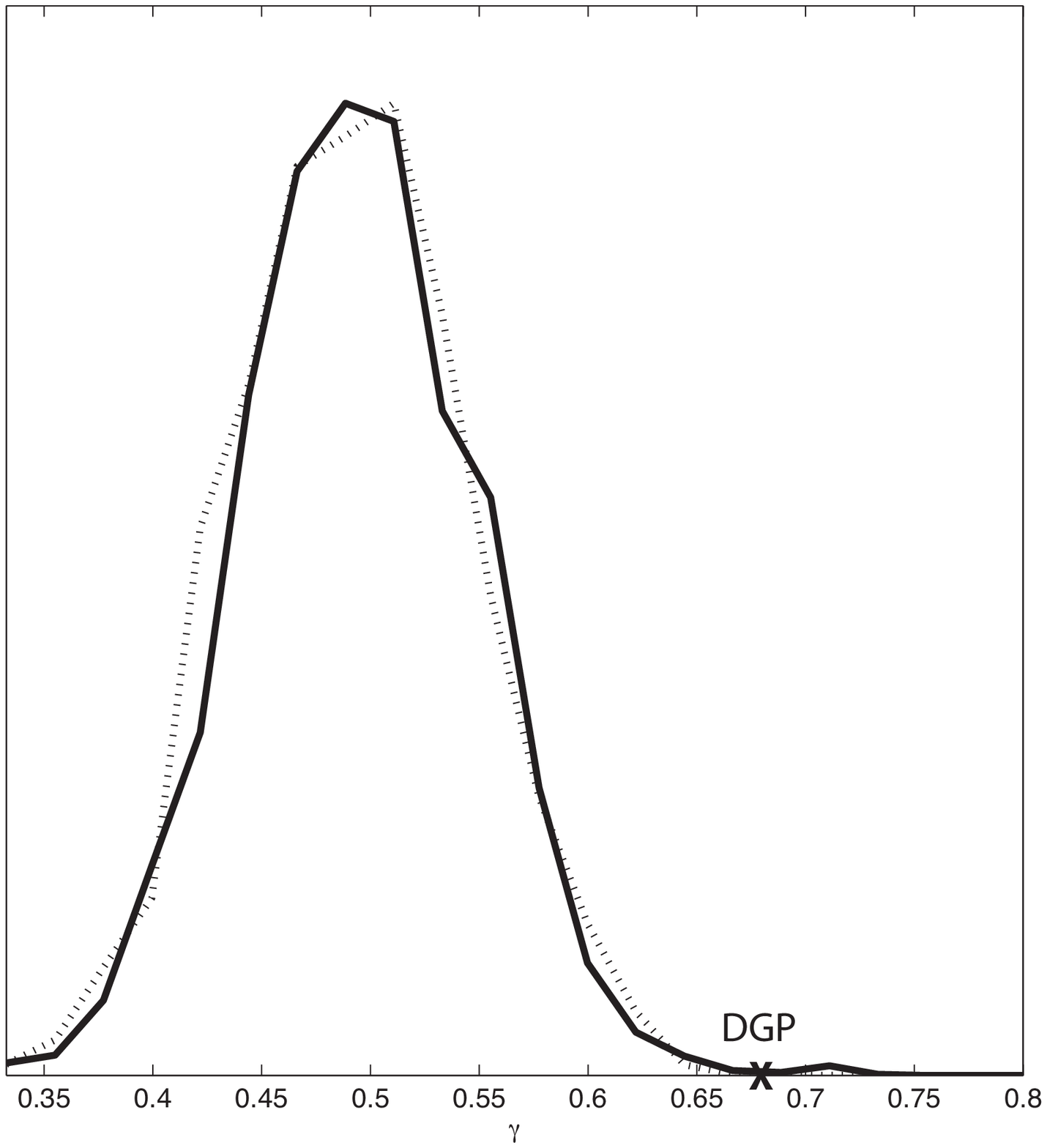}} \\
\hline
\end{tabular}
\caption{\label{fig:G}
Constant Growth Index $\gamma$. TOP LEFT: Moderate scenario fitting background DGP data to an assumed $\Lambda$CDM theoretical model: $\chi^2/dof = 1.063$.  TOP RIGHT: Pessimistic scenario fitting background DGP data on an assumed $\Lambda$CDM theoretical model: $\chi^2/dof = 1.041$.  BOTTOM LEFT:   Moderate scenario fitting background $\Lambda$CDM data to an assumed DGP theoretical model: $\chi^2/dof = 1.051$.  BOTTOM RIGHT: Pessimistic scenario fitting background $\Lambda$CDM data to an assumed DGP theoretical model: $\chi^2/dof = 1.024$.  
}
\end{center}
\end{figure}

Our results are summarized in Table \ref{table:result1}, Figures \ref{fig:INT}, \ref{fig:Exp} and \ref{fig:G}. We find that in either the pessimistic or moderate scenario for future data uncertainties, the incorrect fitted model is ruled out using either of the two parameterizations discussed. We also find that a constant growth index parameter should be able to rule out an incorrect model in the moderate scenario but fails in the pessimistic scenario to rule out $\Lambda$CDM when the DGP data simulated data is used. In addition to the values of $\gamma_0$ and $\gamma_\infty$, the slopes of the interpolated or exponential parameterizations (i.e. $\gamma_a$ and $\gamma_b$) by themselves also rule out an incorrect background.   

\begin{center}
\begin{table}[t]
\label{table:result1}
\begin{tabular}{|c|c|c|c|c|c|c|c|}\hline
\multicolumn{8}{|c|}{\bfseries Interpolated parameterization for the growth index, Eq. (\ref{eq:GammaZ}).}\\ \hline
\multicolumn{1}{|c|}{Background}&\multicolumn{1}{|c|}{Fitted}&
\multicolumn{3}{|c|}{Moderate}&\multicolumn{3}{|c|}{Pessimistic}\\ 
\multicolumn{1}{|c|}{Model}&\multicolumn{1}{|c|}{Theory}&
\multicolumn{3}{|c|}{Scenario}&\multicolumn{3}{|c|}{Scenario}\\ \hline
& &$\mathbf{\gamma_0}$&$\mathbf{\gamma_a}$&$\chi^2/dof$&$\mathbf{\gamma_0}$&$\mathbf{\gamma_a}$&$\chi^2/dof$\\ \hline 
DGP & $\Lambda$CDM &$0.483_{-0.291}^{+0.264}$&$2.26_{-1.05}^{+1.38}$&$1.038$&$0.191_{-0.266}^{+0.432}$&$2.32_{-1.64}^{+1.29}$&$1.023$\\ \hline
$\Lambda$CDM & DGP &$0.727_{-0.307}^{+0.355}$&$-1.62_{-0.95}^{+1.04}$&$1.037$&$0.716_{-0.455}^{+0.457}$&$-1.43_{-1.38}^{+1.52}$&$1.017$\\ \hline 
\hline
\multicolumn{8}{|c|}{\bfseries Exponential parameterization for the growth index, Eq. (\ref{eq:gamexp}).}\\ \hline
\multicolumn{1}{|c|}{Background}&\multicolumn{1}{|c|}{Fitted}&
\multicolumn{3}{|c|}{Moderate}&\multicolumn{3}{|c|}{Pessimistic}\\ 
\multicolumn{1}{|c|}{Model}&\multicolumn{1}{|c|}{Theory}&
\multicolumn{3}{|c|}{Scenario}&\multicolumn{3}{|c|}{Scenario}\\ \hline
& &$\mathbf{\gamma_\infty}$&$\mathbf{\gamma_b}$&$\chi^2/dof$&$\mathbf{\gamma_\infty}$&$\mathbf{\gamma_b}$&$\chi^2/dof$\\ \hline
DGP & $\Lambda$CDM &$1.24_{-0.27}^{+0.25}$&$-0.730_{-0.344}^{+0.388}$&$1.032$&$1.04_{-0.30}^{+0.35}$&$-0.631_{-0.501}^{+0.425}$&$1.022$\\ \hline 
$\Lambda$CDM & DGP &$0.224_{-0.148}^{+0.180}$&$0.475_{-0.345}^{+0.308}$&$1.032$&$0.304_{-0.220}^{+0.255}$&$0.333_{-0.379}^{+0.402}$&$1.015$\\ \hline \hline
\multicolumn{8}{|c|}{\bfseries Constant growth index, $\gamma$.}\\ \hline
\multicolumn{1}{|c|}{Background}&\multicolumn{1}{|c|}{Fitted}&
\multicolumn{3}{|c|}{Moderate}&\multicolumn{3}{|c|}{Pessimistic}\\ 
\multicolumn{1}{|c|}{Model}&\multicolumn{1}{|c|}{Theory}&
\multicolumn{3}{|c|}{Scenario}&\multicolumn{3}{|c|}{Scenario}\\ \hline
& &\multicolumn{2}{|c|}{$\mathbf{\gamma}$}&$\chi^2/dof$&\multicolumn{2}{|c|}{$\mathbf{\gamma}$}&$\chi^2/dof$\\ \hline
DGP & $\Lambda$CDM &\multicolumn{2}{|c|}{$0.818_{-0.099}^{+0.101}$}&$1.063$&\multicolumn{2}{|c|}{$0.645_{-0.102}^{+0.116}$}&$1.041$\\ \hline 
$\Lambda$CDM & DGP &\multicolumn{2}{|c|}{$0.461_{-0.063}^{+0.091}$}&$1.051$&\multicolumn{2}{|c|}{$0.483_{-0.088}^{+0.113}$}&$1.024$\\ \hline 
\end{tabular}
\caption{Here we summarize our best fits for multiple parameterizations of the growth index, when we fit generated background data to the wrong model as described above.  We see that all of the parameterizations are able to find inconsistencies in at least one parameter compared to their expected theoretical values given right below.  In these fits we find best fit values of $\Omega_m^0 \simeq 0.230$ and $H_0 \simeq 74.3$ when the background model is DGP and the fitted theoretical model is $\Lambda$CDM; and $\Omega_m^0 \simeq 0.295$ and $H_0 \simeq 63.4$ when the background is $\Lambda$CDM and the fitted theoretical model is DGP. For these values of $\Omega_m^0$ we would expect the following parameter values for the fitted theoretical models: $(0.564,-0.025)$ for $\Lambda$CDM and $(0.655, 0.046)$ for DGP in the interpolation parameterization; $(0.546,0.011)$ for $\Lambda$CDM and $(0.688, -0.020)$ for DGP in the exponential parameterization; and $(0.552)$ for $\Lambda$CDM and $(0.680)$ for DGP with a constant $\gamma$.  
}
\end{table}
\end{center}
\section{Conclusion}
We explored comparisons of redshift parameterizations of the growth factor index to current and future growth data. The first parametrization used was introduced in previous work and interpolates between a redshift dependent form at small redshifts and a constant value at high redshifts. A second parametrization based on an exponential form is introduced here and exhibits a similar redshift dependence, as it should. We found it to fit theoretical data to within $0.015\%$ for $\Lambda$CDM and $0.09\%$ for DGP, over the entire redshift range up to the CMB surface. While more precise parametrizations are welcome, we consider that the more 
significant plus in these redshift dependent parametrizations is that they provide the slope of the growth index as a second test to the underlying gravity model. This is relevant because the slope is related to variations in $\gamma(z)$ at small redshifts where more data can be obtained. 
Using redshift dependent parameterizations or constant value of the growth index, we find that current growth data from redshift distortions and Lyman alpha forests is unable to put significant constraints on the growth parameters. In order to explore how well future growth data could constrain these parameters, we  simulated growth data and ran a Monte-Carlo-Markov-Chain analysis. We find that a pessimistic or moderate scenarios for future data uncertainties will be able to rule out an incorrectly assumed theoretical model using any of the two parameterizations discussed while we find that in our pessimistic scenario a constant growth index parameter will be unable to rule out an incorrect model. This is due to the fact that the slope acts as a second discriminator at smaller redshifts.
\acknowledgements
We thank L. Guzzo, M. Viel, J. Xia, D.Eisenstein and W. Percival for useful comments about future BAO constraints, and C. Allison for help with graphics. M.I. acknowledges that this material is based upon work supported in part by NASA under grant NNX09AJ55G and that part of the calculations for this work have been performed on the Cosmology Computer Cluster funded by the Hoblitzelle Foundation.

\end{document}